\documentclass[11pt,a4paper]{article}

\usepackage{cite}
\usepackage{graphicx}
\usepackage{graphics}
\usepackage{epsfig}
\usepackage{subfigure}
\usepackage{amsmath}
\usepackage{float}
\usepackage{afterpage}
\usepackage{mathrsfs}
\usepackage{axodraw}

\begin{document}

\begin{titlepage}
\setlength{\topmargin}{1.5cm}

\begin{flushright}
MAN/HEP/2008/3\\
CERN-PH-TH/2008-044
\end{flushright}
\vspace{\fill}

\begin{center}

{\Large {\bf TeVJet: A general framework for the calculation of jet observables in NLO QCD}}

\vspace{1cm}

{\large Michael H. Seymour$^{ab}$, Christopher Tevlin$^a$}

\vspace{0.5cm}

{\large $a$ {\it School of Physics \& Astronomy, University of Manchester, U.K.}}\\
{\large $b$ {\it Theoretical Physics Group, CERN, CH-1211 Geneva 23, Switzerland.}}

\vspace{\fill}

\begin{abstract}
In this paper we present the parton level Monte Carlo program \texttt{TeVJet}, a direct implementation of the dipole subtraction method for calculating jet cross sections in NLO QCD. It has been written so as to allow the inclusion of new processes in as straightforward a way as possible. The user must provide the usual ingredients for an NLO calculation and from these the process-independent parts required to make the phase space integrals finite in 4 dimensions are automatically generated. These integrals are then performed using Monte Carlo techniques. We present the results for a few example processes.
\end{abstract}
\vspace{\fill}

\vspace{\fill}

\vspace{\fill}

\end{center}
\end{titlepage}

\tableofcontents
\newpage
\section{Introduction}
Since the equations of motion derived from the QCD Lagrangian are non-linear coupled equations, they cannot be solved exactly. At large momentum transfers we can exploit the property of asymptotic freedom and use perturbation theory to expand about the partonic states of quarks and gluons and calculate partonic cross sections. In the case of hadron colliders one must then fold these with non-perturbative but universal parton distribution functions. One of the most powerful tests of QCD and our understanding of the strong interaction has been the comparison between experimental data and such perturbative calculations.

Leading order (LO) calculations can be useful in interpreting the experimental data, and there exist several programs that have automated the calculation of observables to this order (some examples are given in Refs.~\cite{Maltoni:2002qb,Mangano:2002ea,Krauss:2001iv}). Higher order terms in perturbation theory have a richer structure, indeed it is not until one calculates the 1-loop correction that we encounter ultraviolet divergences and need to renormalize the field theory. Leading order calculations can only, for some scattering processes, describe the shapes of distributions well but do not say much about the overall normalization. This manifests itself as a strong dependence on the unphysical renormalization and factorization scales.

Thus the calculation of observables to next to leading order (NLO) accuracy is more involved, which, at a practical level, is due to the presence of singularities at intermediate steps in the calculation. This includes both the ultraviolet divergences already mentioned, which require the machinery of renormalization to tame; and infrared divergences, which are a consequence of a field theory with massless quanta. These infrared singularities pose a significant practical challenge in the calculation of exclusive observables, i.e.\ when the available phase space is restricted. The problem is that the cancellation occurs between two quantities that result from integration over phase space volumes of different dimensionality, which makes the numerical integration over these phase space volumes non-trivial~-- each of these integrals is divergent (although their sum is finite).

There are basically two methods for overcoming these problems and performing a (semi-) numerical calculation of these phase space integrals: the phase space slicing method and the subtraction method. Both of these exploit the universality of the singular behaviour of tree level QCD amplitudes in the soft and collinear limits. One of the advantages of such numerical calculations is that several different observables (jet functions) can be calculated in parallel. 

The earliest NLO calculation of an event shape distribution\cite{Ellis:1980wv} introduced and used the subtraction method.  In Ref.~\cite{Kunszt:1989km}, this was developed to produce the first general-purpose program, designed to calculate the NLO corrections for many observables simultaneously, for a single process, $e^+e^-$ annihilation.  Other early calculations used the phase space slicing method, with varying degrees of success\cite{Kramer:1986sg}.  Refs.~\cite{Giele:1991vf,Giele:1993dj} defined the first general method for calculating arbitrary infrared-safe observables in arbitrary processes, using the phase space slicing method.  Refs.~\cite{Catani:1996jh,Catani:1996vz} did the same thing for the subtraction method, defining the dipole subtraction method, which has been the basis for the majority of numerical NLO calculations performed since then.  Ref.~\cite{Catani:1996vz} also succeeded in solving the analogous, but more general, problem with an arbitrary number of identified particles in the final state.  This was also done for the phase-space slicing method in Ref.~\cite{Keller:1998tf}. More recently, those seeking to define a subtraction algorithm for NNLO calculations (for examples, see Refs.~\cite{GehrmannDeRidder:2005cm,Somogyi:2006da,Somogyi:2006db}) have been lead to define new variations on the dipole subtraction theme.  The generalization to massive partons was done for the phase-space slicing method in Ref.~\cite{Keller:1998tf} and the dipole subtraction method in Ref.~\cite{Catani:2002hc}.

In this publication we describe the parton level Monte Carlo program \texttt{TeVJet} which is a direct implementation of the dipole subtraction method, at present in its massless version without identified particles in the final state~\cite{Catani:1996vz}. In section~\ref{dip} we give a brief overview of the dipole subtraction method; in section~\ref{des} we describe in detail the structure of the code; in section~\ref{use} we describe how to use the program with a simple example main program and in section~\ref{result} we present some results for the test processes $e^+e^-\to 3$ jets and $ep\to 1$ jet.

\section{The Dipole Subtraction Method}
\label{dip}
\subsection{Overview of the method}
In this section we introduce the basic idea of the dipole subtraction method~\cite{Catani:1996vz} for the simple case of processes with no hadrons in the initial state (i.e.\ $e^+e^-$ annihilation processes). In section~\ref{hadron} we shall generalize this to include processes with hadron beams. Finally in sections~\ref{fac}--\ref{end} we shall discuss some of the details that are useful for an understanding of \texttt{TeVJet}. 

At next-to-leading order (NLO) in perturbative QCD some observable, $\sigma$, is given by
\begin{equation}
\sigma=\sigma^{\rm LO}+\sigma^{\rm NLO},
\end{equation} 
where $\sigma^{\rm LO}$ is given by the integration of the Born level cross section over the available phase space, i.e.
\begin{eqnarray}
\sigma^{\rm LO}&=&\int_{m}d\sigma^{\rm B}.
\end{eqnarray}
The subscript signifies that this is an integral over $m$-body phase space so explicitly:
\begin{eqnarray}
\sigma^{\rm LO}&=&\frac{1}{\mathcal{F}}\mathcal{N}_{\rm in}\sum_{\{m\}}\frac{1}{S_{\{m\}}}\int d\phi^{(m)}|\mathcal{M}^B_m|^2F_J^{(m)},
\label{born}
\end{eqnarray}
where $\mathcal{F}$ is the flux factor, $\mathcal{N}_{\rm in}$ represents all of the QCD independent factors, $\sum_{\{m\}}$ denotes the sum over all $m$-parton configurations, $S_{\{m\}}$ is the Bose symmetry factor, $d\phi^{(m)}$ is an element of $m$-body phase space, $\mathcal{M}^B_m$ is the matrix element in the Born approximation, and $F_J^{(m)}$ is the jet function, which defines the observable to be calculated. The jet function warrants a little explanation~-- it is the value of the observable as a function of the momenta of the $m$ final state particles. If one wishes to calculate a total cross section this could be simply given by $F_J=1$, or for a total cross section with some cuts it could be some combination of theta functions such that it is given by $F_J=1$ for all kinematic configurations satisfying the cut, and  $F_J=0$ for all other configurations. For a differential cross section it will contain a delta function. Alternatively it could simply contain an analytical function of the final state momenta. In general it will contain a combination of these possibilities but there are some constraints on it that we shall come back to later. Note that by definition the leading order contribution, $\sigma^{\rm LO}$, is finite and the phase space integration can be carried out in four dimensions. This implies that either $\mathcal{M}^B_m$ refers to an intrinsically infrared safe process, or the jet function vanishes ($F_J^{(m)}\rightarrow 0$) in all soft and collinear limits ($p_i.p_j\rightarrow 0$).

The next-to-leading order contribution, $\sigma^{\rm NLO}$, is given by the sum of two parts~-- the {\it real} part, which is the contribution from the real emission Feynman diagrams, and the {\it virtual} part, which is the interference between the one-loop and Born level amplitudes:
\begin{eqnarray}
\sigma^{{\rm NLO}}&=&\int_{m+1}d\sigma^{R}+\int_{m}d\sigma^{V}\\
&=&\int d\phi^{(m+1)}|\mathcal{M}^{R}|^2F_J^{(m+1)}+\int d\phi^{(m)}|\mathcal{M}^{V}|^2F_J^{(m)},
\end{eqnarray}
where $\mathcal{M}^{R}$ is the real matrix element, and by slight abuse of notation we write the interference term as $|\mathcal{M}^{V}|^2$. These are separately divergent in four dimensions although their sum is finite. In fact the sum of these two terms is only finite for infrared safe observables, defined as an observable whose value is independent of the number of soft and collinear hadrons (partons) in the final state. This means that we require the following property of the jet function:
\begin{equation}
F_J^{(m+1)}\rightarrow F_J^{(m)},
\end{equation}
as one approaches an $m+1$-parton configuration that is kinematically degenerate with an $m$-parton configuration.

One way to proceed is to regularize these integrals, e.g.\ by dimensional regularization, which would lead to single ($1/\epsilon$) and double ($1/\epsilon^2$) poles, and then calculate the integrals analytically. Once the two are summed the $\epsilon$ poles will cancel and the limit $\epsilon\rightarrow 0$ can be taken. However we would like to calculate the integrals numerically using Monte Carlo techniques, so we need another strategy.

The behaviour of tree level QCD amplitudes in the soft and collinear limits is well known. The function $|\mathcal{M}^{R}|^2$ is a complicated function that diverges in several regions of phase space, whenever a gluon becomes arbitrarily soft, or when any pair of massless partons becomes collinear. The basic idea of the dipole subtraction algorithm is to construct a local counter term for the real contribution:
\begin{eqnarray}
\sigma^{\rm NLO}=\int_{m+1}[d\sigma^R-d\sigma^A]+\int_{m+1}d\sigma^A+\int_md\sigma^V,
\label{a}
\end{eqnarray}
such that the first term on the RHS of Eq.~(\ref{a}) is finite. The counter term, $d\sigma^A$, is constructed as the sum of several functions each of which diverges in fewer places than $d\sigma^R$. In fact it is constructed, using an improved factorization, from the Born level cross section:
\begin{eqnarray}
d\sigma^A=\sum_{\rm{dipoles}}d\sigma^B\otimes dV_{\rm{dipole}},
\label{dipolefac}
\end{eqnarray}
where by $d\sigma^B$ we mean an appropriate colour + spin projection of the Born level cross section; and the dipole term, $dV_{\rm{dipole}}$, represents a $2\rightarrow 3$ splitting process.

One should think of these terms as being produced in a two step process: first an $m$-parton configuration with Born-like kinematics is produced; then one of these particles (the emitter) decays into two particles in the presence of another (the spectator). The dipole terms are process independent and universal and only depend on the quantum numbers of the emitter and spectator. As we shall see in the practical implementation, the construction of the counter term is the reverse of this process~-- first an $m+1$-parton configuration is produced and then this is mapped onto an $m$-parton configuration.


Of course the remaining two integrals on the RHS of Eq.~(\ref{a}) must be combined for the method to work, and we shall sketch out how this works now. In Eq.~(\ref{dipolefac}) the subtraction cross section, $d\sigma^A$, is evaluated at the $m+1$-parton phase space point $(p_1,\dots,p_{m+1})$. It is constructed by choosing three of the $m+1$ partons, $i,j,k$ (summing over all possible such choices), and projecting their momenta, $p_i,p_j,p_k$ onto new momenta, which we call $\widetilde{p}_{ij},\widetilde{p}_k$, and all other momenta $\{p_l\}$ onto $\{\widetilde{p}_l\}$. That is, formally, the tilde momenta, $\{\widetilde{p}\}$, are functions of the momenta $\{p\}$ and of which partons are involved in the splitting process, i.e.\ $i$, $j$ and $k$.

We describe $\widetilde{ij}$ and $\widetilde{k}$ as the emitter and spectator respectively. Their momenta are chosen to be linear combinations of $p_i,p_j,p_k$ such that: they are on mass-shell ($\widetilde{p}_{ij}^2=\widetilde{p}_{k}^2=0$); momentum is conserved ($p_i^{\mu}+p_j^{\mu}+p_k^{\mu}=\widetilde{p}_{ij}^{\mu}+\widetilde{p}_k^{\mu}$); and the phase space of the partons $i,j,k$ can be factorized into the dipole phase space times a single parton phase space:
\begin{eqnarray}
d\phi^{(3)}(p_i,p_j,p_k;Q)&=&d\phi^{(2)}(\widetilde{p}_{ij},\widetilde{p}_k;Q)\left[dp_i(\widetilde{p}_{ij},\widetilde{p}_k)\right]\nonumber\\
&=&\mathcal{J}d\phi^{(2)}(\widetilde{p}_{ij},\widetilde{p}_k;Q)d\Omega^{(d-3)}dz\,dy.
\label{phase-fact}
\end{eqnarray}
Here $Q$ is the total momentum of the 3-parton system, $\mathcal{J}$ is a Jacobian factor for the mapping, and $\Omega^{(d-3)}$, $z$ and $y$ are the degrees of freedom of the $(d-1)$-dimensional single parton subspace. In particular, $\Omega^{(d-3)}$ represents a set of $d-3$ angles that define the direction of parton $i$ in the transverse subspace: in four dimensions it reduces to the azimuthal angle $\phi$. All other momenta remain unchanged. The second term on the RHS of Eq.~(\ref{a}) is given by:
\begin{equation}
\int_{m+1}d\sigma^A=\int d\phi^{(m+1)}(p;Q)\sum_{\rm{dipoles}}\frac{d\sigma^B(\widetilde{p})}{d\phi^{(m)}(\widetilde{p})}\otimes\frac{dV_{\rm{dipole}}(p)}{\left[dp_i(\widetilde{p}_{ij},\widetilde{p}_k)\right]}.
\label{subtle1}
\end{equation}
Since the three parton phase space factorizes and the dipole term, $dV_{\rm{dipole}}/dp_i$, is the only term that depends on the single parton phase space we can write\footnote{Note a slight subtlety in going from Eq.~(\ref{subtle1}) to Eq.~(\ref{subtle3}): Exchanging the order of summing and integrating has changed the meaning of the momenta: In Eq.~(\ref{subtle1}), each dipole defines a different set of tilde momenta $\{\widetilde{p}\}$ for the same momenta $\{p\}$, whereas in Eq.~(\ref{subtle2}), each dipole defines a different set of momenta $\{p\}$ for the same tilde momenta $\{\widetilde{p}\}$. Since the momenta are dummy variables of integration and the exact phase space factorization has the same form for each dipole, this is a completely transparent change. Having made it, the order can be re-exchanged as in Eq.~(\ref{subtle3}).}
\begin{eqnarray}
\int_{m+1}d\sigma^A&=&\sum_{\rm{dipoles}}\int d\phi^{(m)}(\widetilde{p};Q)\frac{d\sigma^B(\widetilde{p})}{d\phi^{(m)}(\widetilde{p})}\otimes\int d\Omega^{(d-3)}dz\,dy\,\mathcal{J}\frac{dV_{\rm{dipole}}(\widetilde{p};\Omega^{(d-3)},z,y)}{\left[dp_i(\widetilde{p}_{ij},\widetilde{p}_k)\right]}\label{subtle2} \phantom{(99)}\\
&=&\int d\phi^{(m)}(\widetilde{p};Q)\frac{d\sigma^B(\widetilde{p})}{d\phi^{(m)}(\widetilde{p})}\otimes{\bf I}(\widetilde{p};\epsilon),
\label{subtle3}
\end{eqnarray}
where the insertion operator, ${\bf I}(\widetilde{p};\epsilon)$, is given by:
\begin{equation}
{\bf I}(\widetilde{p};\epsilon)=\sum_{\rm{dipoles}}\int d\Omega^{(d-3)}dz\,dy\mathcal{J}\frac{dV_{\rm{dipole}}(\widetilde{p};\Omega^{(d-3)},z,y)}{\left[dp_i(\widetilde{p}_{ij},\widetilde{p}_k)\right]}.
\end{equation}
The insertion operator contains all of the divergent behaviour of $d\sigma^A$ and can be calculated analytically in $d$ dimensions. Since the dipole term, $V_{\rm{dipole}}$, and the Jacobian factor, $\mathcal{J}$, are process independent this can be calculated once and for all. 

This can now be combined with the virtual term as it is an integral over $m$-parton phase space (see section~\ref{end}). The NLO contribution is therefore given by:
\begin{eqnarray}
\sigma^{\rm NLO}&=&\sigma^{{\rm NLO}\{m+1\}}+\sigma^{{\rm NLO}\{m\}}\nonumber\\
&=&\int_{m+1}[d\sigma^R-d\sigma^A]+\int_m[d\sigma^V+d\sigma^B\otimes{\bf I}].
\end{eqnarray}
Now we have two phase space integrals, which are finite in 4 dimensions and can be calculated by Monte Carlo integration.

\subsection{Initial state hadrons}
\label{hadron}
When there are hadrons in the initial state, the perturbatively calculable partonic cross section must be folded with non-perturbative, but universal, parton distribution functions. That is the cross section for a reaction with two hadrons in the initial state with momenta $p_1$ and $p_2$ is given by:
\begin{eqnarray}
\sigma(p_1,p_2)&=&\sum_{a,b}\int^{1}_{0}d\eta_a\,f_{h/a}(\eta_a,\mu_{F}^2)\int^{1}_{0}d\eta_b\,f_{h/b}(\eta_b,\mu_{F}^2)\nonumber\\
&&\times[\sigma^{{\rm LO}}_{ab}(\eta_ap_1,\eta_bp_2)+\sigma^{{\rm NLO}}_{ab}(\eta_ap_1,\eta_bp_2,\mu_{F}^2)],
\end{eqnarray}
where the sum is over parton flavours; the parton distribution function (pdf), $f_{h/a}(\eta_a,\mu_{F}^2)$, gives the probability of finding a parton type $a$ inside the hadron with a fraction $\eta_a$ of the momentum of the hadron at some scale $\mu_F$; $\sigma^{{\rm LO}}_{ab}$ and $\sigma^{{\rm NLO}}_{ab}$ are the partonic cross sections, which we shall define below.

The leading order contribution is given by:
\begin{equation}
\sigma^{\rm LO}_{ab}=\int\frac{1}{n_cn_s\mathcal{F}}\mathcal{N}_{\rm in}\sum_{\{m\}}\frac{1}{S_{\{m\}}}d\phi^{(m)}|\mathcal{M}^B_{ab}|^2(\eta_ap_1,\eta_bp_2)F_J^{(m)},
\label{bornloop}
\end{equation}
where $n_c$ and $n_s$ account for the average over the number of initial state spin and colour states, $\mathcal{F}$ is the flux factor, $\mathcal{N}_{\rm in}$ accounts for all QCD independent factors, $\mathcal{M}^B_{ab}$ is the amplitude, in the Born approximation, for partons type $a$ and $b$ to give an $m$-parton final state that contributes to the observable; the sum $\sum_{\{m\}}$ runs over all such final states.

The presence of hadrons (and hence partons) with well defined momenta in the initial state spoils the cancellation of divergences between the real and virtual contributions. The parts in the real contribution that do not cancel are the initial-state collinear divergences, i.e.\ when a final-state parton becomes collinear with one of the initial-state partons. The problem is in fact one of double counting: this is exactly the physics that is already included in the pdfs; and the solution is to subtract off these collinear poles from the real term, which introduces a scale dependence into both the pdfs and the cross section. The NLO contribution is therefore given by:
\begin{equation}
\sigma^{{\rm NLO}}_{ab}(p_1,p_2,\mu_F^2)=\int_{m+1}d\sigma^R_{ab}(p_1,p_2)+\int_{m}d\sigma^V_{ab}(p_1,p_2)+\int_{m}d\sigma^C_{ab}(p_1,p_2,\mu_F^2),
\label{nloloop}
\end{equation}
where the last term is the collinear subtraction term. Although the pole must be subtracted, there is a free choice of how much finite physics is also subtracted and absorbed into the pdf. A particular choice of this sets the factorization scheme and by changing the scheme we simply move physics between the pdf and the perturbative calculation. The collinear counter-term is given by:
\begin{eqnarray}
d\sigma^C_{ab}(p_1,p_2,\mu_F^2)&=&-\frac{\alpha_S}{2\pi}\frac{1}{\Gamma(1-\epsilon)}\sum_{cd}\int_{0}^{1}dz_1\int_{0}^{1}dz_2\,d\sigma^B_{cd}(z_1p_1,z_2p_2)\nonumber\\
&\times&\left\{\delta_{bd}\delta(1-z_2)\left[-\frac{1}{\epsilon}\left(\frac{4\pi\mu^2}{\mu_F^2}\right)^{\epsilon}P^{ac}(z_1)+K^{ac}_{{\rm FS}}(z_1)\right]\right.\nonumber\\
&+&\left.\delta_{ac}\delta(1-z_1)\left[-\frac{1}{\epsilon}\left(\frac{4\pi\mu^2}{\mu_F^2}\right)^{\epsilon}P^{bd}(z_2)+K^{bd}_{{\rm FS}}(z_2)\right]\right\},
\label{col}
\end{eqnarray} 
where $P^{ab}(z)$ are the (four-dimensional) regularized Altarelli-Parisi splitting functions and the kernels $K^{ab}_{{\rm FS}}(z)$ specify the factorization scheme (setting $K^{ab}_{{\rm FS}}(z)=0$ defines the $\overline{\mathrm{MS}}$ subtraction scheme).

In the case of initial state partons there are additional phase space constraints due to the fact that there are hadrons in the initial state with well defined momenta. These constraints are different depending on whether the emitter and spectator are in the final state or initial state. Therefore there are different dipole contributions:
\begin{eqnarray}
d\sigma^A=\sum_{\rm{dipoles}}d\sigma^B\otimes(dV_{\rm{dipole}}+dV_{\rm{dipole}}^{\prime}),
\end{eqnarray}
where $dV_{\rm{dipole}}$ is the dipole term already discussed for emitter and spectator both in the final-state, and $dV_{\rm{dipole}}^{\prime}$ represents the remaining three dipole terms involving initial state emitters and/or spectators.

As before we consider the analytical integration of each dipole contribution over the one-parton subspace leading to the divergence. For the integration of the new dipoles involving initial state partons the analogue of the phase space factorization is a phase space convolution. The specific form of the convolution depends on the phase space constraints, but in the case of a final state emitter, $ij$, and an initial state spectator, $a$, it has the form:
\begin{eqnarray}
d\phi^{(2)}(p_i,p_j;Q+p_a)&=&\int_0^1dx\,d\phi^{(1)}({\widetilde p}_{ij};Q+xp_a)\left[dp_i(\widetilde{p}_{ij};p_a,x)\right]\nonumber\\
&=&\int_0^1dx\,d\phi^{(1)}({\widetilde p}_{ij};Q+xp_a)\mathcal{J}d\Omega^{(d-3)}dz,
\end{eqnarray}
where here $\mathcal{J}$ is a new Jacobian factor. This dipole contribution is given by:
\begin{eqnarray}
\int d\phi^{(m+1)}(p;Q+p_a)\frac{d\sigma^B}{d\phi^{(m)}}\otimes\frac{dV^{a}_{ij}}{\left[dp_i(\widetilde{p}_{ij};p_a,x)\right]}&=&\int_0^1dx\int d\phi^{(m)}({\widetilde p};Q+xp_a)\frac{d\sigma^B}{d\phi^{(m)}}\nonumber\\
&&\otimes\int d\Omega^{(d-3)}dz\mathcal{J}\frac{dV^{a}_{ij}}{\left[dp_i(\widetilde{p}_{ij};p_a,x)\right]}.
\end{eqnarray}
The last part contains all of the dependence on the single parton subspace variables $\Omega^{(d-3)}$ and $z$, and is process independent. Thus this integral can be carried out once and for all in $d$ dimensions. In a similar way the contributions from the other two types of dipole with initial state emitters and/or spectators can be calculated. There is a part of each of these terms that is proportional to $\delta(1-x)$ and these terms combine with the integral of the final-state dipole, $V_{{\rm dipole}}$, to form the insertion operator, ${\bf I}(\widetilde{p};\epsilon)$. This matches the singular behaviour of the virtual term.

It can be shown that all of the divergences from the remaining terms cancel with those of the collinear counter-term in Eq.~(\ref{col}). Thus all of the $1/\epsilon$ poles cancel and the finite parts can be collected together to give a convolution term of the form:
\begin{equation}
\int^1_0dx\left[\hat{\sigma}_{ab}^{{\rm NLO}\{m\}}(x;xp_a,p_b,\mu_F^2)+\hat{\sigma}_{ab}^{{\rm NLO}\{m\}}(x;p_a,xp_b,\mu_F^2)\right],
\label{conv}
\end{equation}
where
\begin{equation}
\int_0^1dx\,\hat{\sigma}_{ab}^{{\rm NLO}\{m\}}(x;xp_a,p_b,\mu_F^2)=\sum_{a^{\prime}}\int_0^1dx\int_m\left[d\sigma_{a^{\prime}b}(xp_a,p_b)\otimes({\bf P}+{\bf K})^{aa^{\prime}}(x)\right]_{\epsilon=0},
\end{equation}
and analogously for $\hat{\sigma}_{ab}^{{\rm NLO}\{m\}}(x;p_a,xp_b,\mu_F^2)$.
Therefore after the dipole subtraction procedure has been carried out the NLO contribution to the jet observable may be written in the following form:
\begin{eqnarray}
\sigma_{ab}^{{\rm NLO}}(p)&=&\sigma_{ab}^{{\rm NLO}\{m+1\}}(p)+\sigma_{ab}^{{\rm NLO}\{m\}}(p)\nonumber\\
&&+\int^1_0dx\left[\hat{\sigma}_{ab}^{{\rm NLO}\{m\}}(x;xp_a,p_b,\mu_F^2)+\hat{\sigma}_{ab}^{{\rm NLO}\{m\}}(x;p_a,xp_b,\mu_F^2)\right]
\end{eqnarray}
where each of the three terms is finite in four dimensions and can be calculated by Monte Carlo integration.
\subsection{Factorization in the soft and collinear limits}
\label{fac}
In order to construct the dipole subtraction terms, we need to know the behaviour of tree-level QCD amplitudes in the divergent regions of phase space, i.e.\ in the soft and collinear limits. In this section we recall these properties of tree-level QCD amplitudes, and in section~\ref{dipole} we shall write down the dipole terms. In order to do both of these it is necessary to introduce the notation we shall use\cite{Catani:1996vz}.

The tree-level matrix element for a scattering process with QCD partons in the initial-state and $m$ QCD partons in the final-state is\footnote{In the following we shall use $i,j,\dots$ to represent final-state partons, and $a,\dots$ to represent initial-state partons.}
\begin{equation}
\mathcal{M}^{c_1,\dots,c_m,c_a,\dots;s_1,\dots,s_m,s_a,\dots}_{m;a,\dots}(p_1,\dots,p_m;p_a,\dots),
\end{equation}
where $\{c_1,\dots,c_m,c_a,\dots\}$, $\{s_1,\dots,s_m,s_a,\dots\}$ and $\{p_1,\dots,p_m,p_a,\dots\}$ label respectively the colour, spin and momenta of the external partons. We introduce a basis $\{|c_1,\dots,c_m,c_a,\dots\rangle\otimes|s_1,\dots,s_m,s_a,\dots\rangle\}$ in helicity $+$ colour space in such a way that
\begin{eqnarray}
\lefteqn{\mathcal{M}^{c_1,\dots,c_m,c_a,\dots;s_1,\dots,s_m,s_a,\dots}_{m;a,\dots}(p_1,\dots,p_m;p_a,\dots)=\sqrt{n_c(a)\dots}}\nonumber\\
&\times\left(\langle c_1,\dots,c_m,c_a,\dots|\otimes\langle s_1,\dots,s_m,s_a,\dots|\right)|1,\dots,m;a,\dots\rangle_{m},
\end{eqnarray}
where $n_c(a)$ is the number of colour states of the initial state parton $a$ ($N_c$ for a quark/anti-quark, $N_c^2-1$ for a gluon). That is $|1,\dots,m;a,\dots\rangle_{m}$ is a vector in colour $+$ helicity space. In this notation the matrix element squared summed(averaged) over final(initial)-state colours is given by:
\begin{equation}
|\mathcal{M}_{m;a,\dots}|^2=_{m;a,\dots}\langle1,\dots,m;a,\dots|1,\dots,m;a,\dots\rangle_{m;a,\dots}.
\end{equation}
In order to describe colour correlations it is convenient also to define a colour charge operator, ${\bf T}_i$, associated with the emission of a gluon from each parton $i$. If the emitted gluon has colour index $c$, the colour charge operator is
\begin{equation}
{\bf T}_i=T^c_i|c\rangle
\end{equation}
and its action onto the colour space is defined by
\begin{eqnarray}
\lefteqn{\langle c_1,\dots,c_i,\dots,c_m,c_a,\dots,c|{\bf T}_i|b_1,\dots,b_i,\dots,b_m,b_a,\dots\rangle}\nonumber\\
&=&\delta_{c_1b_1}\dots T^c_{c_ib_i}\dots\delta_{c_mb_m}\delta_{c_ab_a}\dots\, ,
\end{eqnarray}
where $T^a_{bc}=if_{abc}$ if the emitting particle $i$ is a gluon and $T^a_{\alpha\beta}=t^a_{\alpha\beta}$ if the emitting particle is a quark (in the case of an emitting anti-quark $T^a_{\alpha\beta}=\bar{t}^a_{\alpha\beta}=-t^a_{\beta\alpha}$). In this notation conservation of colour is
\begin{equation}
\left(\sum_{i=1}^{m}{\bf T}_i+{\bf T}_a+\dots\right)|1,\dots,m;a,\dots\rangle_{m}=0.
\end{equation}
Finally, in this notation the square of the colour-correlated matrix element needed for the soft limit is given by
\begin{equation}
|\mathcal{M}^{I,J}_{m;a,\dots}|^2=_{m;a,\dots}\langle1,\dots,m;a,\dots|{\bf T}_I.{\bf T}_J|1,\dots,m;a,\dots\rangle_{m;a,\dots},
\end{equation}
where the indices $I,J$ refer either to final-state or initial-state partons. 
\subsubsection{Soft limit}
Consider the behaviour of a QCD matrix element with $m+1$ final-state partons when one final-state gluon becomes arbitrarily soft. If the momentum of this gluon is $p_j$, the soft limit can be parameterized by:
\begin{eqnarray}
p_j^{\mu}=\lambda q^{\mu},&\lambda\rightarrow 0,
\end{eqnarray}
where $q^{\mu}$ is an arbitrary 4-vector. In the soft limit the matrix element squared with $m+1$ partons tends towards the matrix element squared with $m$ partons found by removing one gluon, multiplied by a divergent factor
\begin{eqnarray}
\lefteqn{_{m+1;a,\dots}\langle 1,\dots,m+1;a,\dots|1,\dots,m+1;a,\dots\rangle_{m+1;a,\dots}\rightarrow-\frac{1}{\lambda^2}4\pi\mu^{2\epsilon}\alpha_S}\nonumber\\
&\times\phantom{}_{m;a,\dots}\langle 1,\dots,m+1;a,\dots|\left[{\bf J}^{\mu}(q)\right]^{\dagger}{\bf J}_{\mu}(q)|1,\dots,m+1;a,\dots\rangle_{m;a,\dots}.
\end{eqnarray}
The eikonal current is given by
\begin{equation}
{\bf J}^{\mu}(q)=\sum_I{\bf T}_I\frac{p_I^{\mu}}{p_I.q},
\end{equation}
where the sum is over all final-state and initial-state partons. Putting these together we re-write the soft limit factorization as
\begin{eqnarray}
\lefteqn{_{m+1;a,\dots}\langle 1,\dots,m+1;a,\dots|1,\dots,m+1;a,\dots\rangle_{m+1;a,\dots}\rightarrow-\frac{1}{\lambda^2}8\pi\mu^{2\epsilon}\alpha_S}\nonumber\\
&\times\sum_I\frac{1}{p_I.q}\sum_{K\ne I}\phantom{}_{m;a,\dots}\langle 1,\dots,m+1;a,\dots|\frac{{\bf T}_K.{\bf T}_Ip_K.p_I}{(p_I+p_K).q}|1,\dots,m+1;a,\dots\rangle_{m;a,\dots}.\nonumber\\
\phantom{a}
\label{soft}
\end{eqnarray}

We point out two properties of the factorization in Eq.~(\ref{soft}) that are relevant for the construction of the dipole subtraction terms. First we note that the Born level matrix element squared does not factorize exactly, for there are colour correlations relating to the fact that an arbitrarily soft gluon cannot resolve the colour charges of individual partons, but is emitted coherently from (at next-to-leading order) all {\it pairs} of partons. It is one of the roles of the spectator parton in the dipole subtraction method to account for these colour correlations. Next we note that this factorization is only valid in the soft limit. If one wished to define a counter term for the soft limit in such a way that the factorization was valid over the whole of phase space some care is needed. In a $1\rightarrow 2$ splitting process, if one wishes to ensure that all external particles are on mass-shell, momentum cannot be conserved. This is the other role of the spectator in the dipole method~-- it absorbs some momentum so that momentum conservation is implemented exactly (see note in section~\ref{initial} about the initial-initial dipole $\mathcal{D}^{ai,b}$).

\subsubsection{Collinear limit}
Now consider the limit in which two final-state QCD partons with momenta $p_i$ and $p_j$ become collinear, parameterized as follows:
\begin{eqnarray}
p_i^{\mu}&=&zp^{\mu}+k_{\perp}^{\mu}-\frac{k_{\perp}^2}{z}\frac{n^{\mu}}{2p.n},\nonumber\\
p_j^{\mu}&=&(1-z)p^{\mu}-k_{\perp}^{\mu}-\frac{k_{\perp}^2}{1-z}\frac{n^{\mu}}{2p.n},\nonumber\\
2p_i.p_j&=&-\frac{k_{\perp}^2}{z(1-z)},\nonumber\\
k_{\perp}&\rightarrow&0,
\end{eqnarray}
where $p^{\mu}$ is a light-like vector ($p^2=0$) that gives the collinear direction; $k^{\mu}_{\perp}$ is a time-like vector transverse to the collinear direction ($k^{2}_{\perp}<0$, $k_{\perp}.p=0$); and $n^{\mu}$ is an additional light-like vector necessary to specify how the collinear limit is approached, which satisfies $k_{\perp}.n=0$. As we approach the collinear limit the matrix element squared for $m+1$ partons factorizes into the matrix element squared for $m$ partons multiplied by a divergent splitting kernel:
\begin{equation}
|\mathcal{M}_{m+1}|^2\rightarrow |\mathcal{M}_m|^2|\mathcal{M}_{(ij)\rightarrow i+j}|^2.
\end{equation}
The $m$ parton matrix element is found by replacing partons $i$ and $j$ with a single parton $ij$ with the addition rule (from QCD vertices) that anything+gluon becomes anything, and quark+antiquark becomes a gluon. In the notation of Ref.~\cite{Catani:1996vz}:
\begin{eqnarray}
\lefteqn{_{m+1;a,\dots}\langle 1,\dots,m+1;a,\dots|1,\dots,m+1;a,\dots\rangle_{m+1;a,\dots}\rightarrow\frac{1}{p_i.p_j}4\pi\mu^{2\epsilon}\alpha_S}\nonumber\\
&\times\phantom{}_{m;a,\dots}\langle 1,\dots,m+1;a,\dots|\hat{P}_{(ij),i}(z,k_{\perp};\epsilon)|1,\dots,m+1;a,\dots\rangle_{m;a,\dots}\,.
\label{collinear}
\end{eqnarray}

The splitting kernel $\hat{P}_{(ij),i}(z,k_{\perp};\epsilon)$ is the unregularized $d$-dimensional Altarelli-Parisi splitting function, note that it depends not only on the collinear momentum fraction, $z$, but also on the transverse momentum $k_{\perp}$. We note that the Born level matrix element does not factorize exactly, in fact there are spin correlations arising from the fact that the splitting kernel is a matrix in the helicity space of the parton $ij$ acting on the helicity indices of parton $ij$ in the vectors $_{m;a,\dots}\langle 1,\dots,m+1;a,\dots|$ and $|1,\dots,m+1;a,\dots\rangle_{m;a,\dots}$. We do not list the splitting functions here, but refer the reader to~\cite{Catani:1996vz}.

Next consider the limit that a final-state parton $i$ becomes collinear with an initial-state parton $a$, parameterized as:
\begin{eqnarray}
p_i^{\mu}&=&(1-x)p^{\mu}_a+k_{\perp}^{\mu}-\frac{k_{\perp}^2}{1-x}\frac{n^{\mu}}{2p_a.n},\nonumber\\
2p_i.p_a&=&-\frac{k_{\perp}^2}{1-x},\nonumber\\
k_{\perp}&\rightarrow&0,
\end{eqnarray}
and the splitting process $a\rightarrow ai+i$ involves a transition from the initial-state parton $a$ to the initial-state parton $ai$ with the associated emission of the final-state parton $i$. Again the quantum numbers of the emitter $ai$ are assigned according to their conservation at the tree-level QCD vertices: if $a$ and $i$ are the same type, the $ai$ is a gluon; if $a$ is a fermion(gluon) and $i$ is a gluon(fermion) then $ai$ will be a fermion(antifermion). The collinear limit as $p_i.p_a\rightarrow 0$ reads
\begin{eqnarray}
\lefteqn{_{m+1;a,\dots}\langle 1,\dots,m+1;a,\dots|1,\dots,m+1;a,\dots\rangle_{m+1;a,\dots}}\nonumber\\
&\rightarrow&\frac{1}{x}\frac{1}{p_i.p_a}4\pi\mu^{2\epsilon}\alpha_S\nonumber\\
&&\times\phantom{}_{m;ai,\dots}\langle 1,\dots,m+1;ai,\dots|\hat{P}_{a,(ai)}(z,k_{\perp};\epsilon)|1,\dots,m+1;ai,\dots\rangle_{m;ai,\dots}\,.
\label{collinear-init}
\end{eqnarray}
The $m$-parton matrix element on the RHS is found from the $m+1$-parton matrix element by removing parton $i$, and replacing parton $a$ with parton $ai$ (note that this parton carries momentum $xp_a$).

As with the soft limit the factorization in Eqs.~(\ref{collinear}) and~(\ref{collinear-init}) is only valid in the collinear limit, and some care must be taken when extrapolating away from the collinear regions of phase space. Finally we note that the soft and collinear limits overlap: it is possible for a final-state gluon to become both soft and collinear with another external parton. When constructing a local counter term for $d\sigma^R$ one must ensure that one avoids double-counting of these singular regions.

\subsection{The dipole contributions}
\label{dipole}
In this section we shall construct the local counter term, $d\sigma^A$, for the real term. It has the property that as we approach the soft and collinear limits it reproduces the factorizations of Eqs.~(\ref{soft}) and~(\ref{collinear}), i.e.\ as we approach a singular area of phase space
\begin{equation}
d\sigma^A\rightarrow d\sigma^R.
\end{equation}
Since momentum is conserved exactly, these limits are approached smoothly thus avoiding double counting in the regions that are both soft and collinear. As well as requiring that $d\sigma^A$ matches the singular behaviour of $d\sigma^R$ we also require that it has no additional singularities~-- a point that we shall return to.

\begin{figure}[t]
\newcommand{\cD}{\mathcal{D}}
\newcommand{\bV}{\mathbf{V}}
\centerline{
\begin{picture}(360,230)(0,0)
\put(0,120){
  \begin{picture}(160,120)(0,0)
  \Line(80,50)(140, 90)
  \Line(80,50)(120, 20)
  \LongArrow( 90,30)(105, 18)
  \LongArrow(115,85)(130, 95)
  \LongArrow(115,55)(130, 45)
  \Line(110,70)(140,50)
  \Vertex(110,70){2.5}
  \GCirc(80,50){10}{1}
  \put( 90, 70){$\widetilde{ij}$}
  \put(145, 88){$i$}
  \put(145, 46){$j$}
  \put(125, 15){$k$}
  \put( 10,110){\underline{$\cD_{ij,k}$}:}
  \put( 40, 50){$\bV_{ij,k}$}
  \put( 90, 14){$p_k$}
  \put(115, 99){$p_i$}
  \put(116, 40){$p_j$}
  \end{picture} }
\put(200,120){
  \begin{picture}(160,120)(0,0)
  \Line(80,50)(140, 90)
  \Line(80,50)( 40, 20)
  \LongArrow( 52,18)( 67, 30)
  \LongArrow(115,85)(130, 95)
  \LongArrow(115,55)(130, 45)
  \Line(110,70)(140,50)
  \Vertex(110,70){2.5}
  \GCirc(80,50){10}{1}
  \put( 90, 70){$\widetilde{ij}$}
  \put(145, 88){$i$}
  \put(145, 46){$j$}
  \put( 28, 18){$a$}
  \put( 10,110){\underline{$\cD_{ij}^a$}:}
  \put( 40, 50){$\bV_{ij}^a$}
  \put( 65, 18){$p_a$}
  \put(115, 99){$p_i$}
  \put(116, 40){$p_j$}
  \end{picture} }
\put(0,-10){
  \begin{picture}(160,120)(0,0)
  \Line(80,50)( 20, 90)
  \Line(80,50)(120, 20)
  \LongArrow( 90,30)(105, 18)
  \LongArrow( 65,70)( 80, 80)
  \LongArrow( 20,80)( 35, 70)
  \Line( 50,70)( 80,90)
  \Vertex( 50,70){2.5}
  \GCirc(80,50){10}{1}
  \put( 49, 46){$\widetilde{ai}$}
  \put( 10, 88){$a$}
  \put( 87, 90){$i$}
  \put(125, 15){$j$}
  \put( 10,110){\underline{$\cD_j^{ai}$}:}
  \put(100, 50){$\bV_j^{ai}$}
  \put( 90, 14){$p_j$}
  \put( 20, 63){$p_a$}
  \put( 76, 67){$p_i$}
  \end{picture} }
\put(200,-10){
  \begin{picture}(160,120)(0,0)
  \Line(80,50)( 20, 90)
  \Line(80,50)( 40, 20)
  \LongArrow( 52,18)( 67, 30)
  \LongArrow( 65,70)( 80, 80)
  \LongArrow( 20,80)( 35, 70)
  \Line( 50,70)( 80,90)
  \Vertex( 50,70){2.5}
  \GCirc(80,50){10}{1}
  \put( 49, 46){$\widetilde{ai}$}
  \put( 10, 88){$a$}
  \put( 87, 90){$i$}
  \put( 28, 18){$b$}
  \put( 10,110){\underline{$\cD^{ai,b}$}:}
  \put(100, 50){$\bV^{ai,b}$}
  \put( 65, 18){$p_b$}
  \put( 20, 63){$p_a$}
  \put( 76, 67){$p_i$}
  \end{picture} }
\end{picture} }
\caption{Illustration of the four dipole types we consider in this paper: final-state emitter with final-state spectator ($\cD_{ij,k}$), final-state emitter with initial-state spectator ($\cD_{ij}^a$), initial-state emitter with final-state spectator ($\cD_j^{ai}$) and initial-state emitter with initial-state spectator ($\cD^{ai,b}$).}
\label{dipoles}
\end{figure}
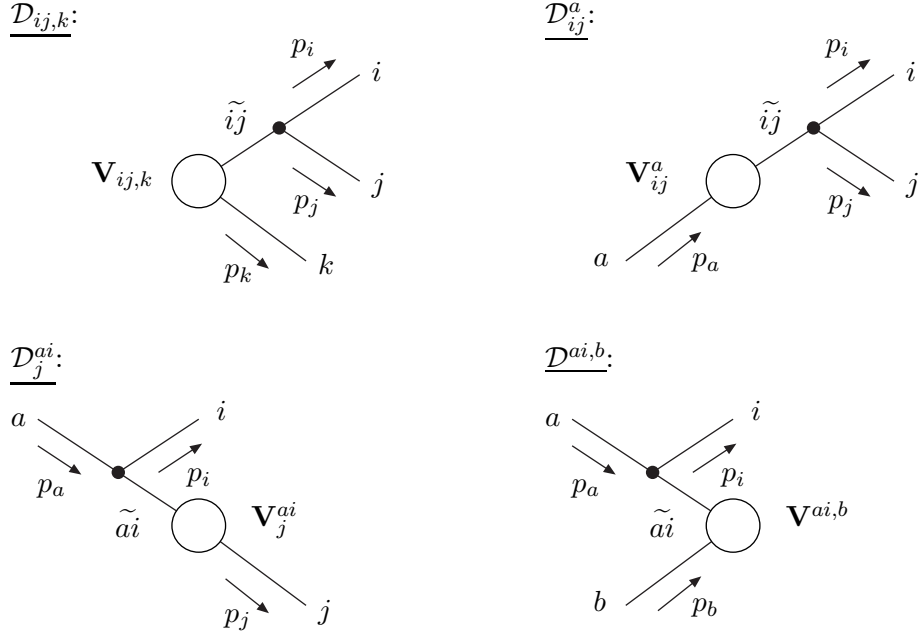

In fact there are several ways of extrapolating smoothly from the singular regions of phase space, and we shall introduce four different counter terms to match the different phase space constraints. In the following we shall use a pair of indices to denote an emitter, and a single index to denote a spectator. These will appear as a subscript (superscript) to denote a final (initial) state parton. Figure~\ref{dipoles} shows the four different types of dipole contribution that act as counter terms for the real matrix element squared, $|\mathcal{M}_{m+1;a,\dots}|^2$, in different regions of phase space.

\subsubsection{Final-state singularities}
Using the dipole factorization formula we can write the real matrix element squared as we approach a final-state singularity ($p_i.p_j\rightarrow 0$) as
\begin{eqnarray}
\lefteqn{_{m+1;a,b}\langle1,\dots,m+1;a,b|1,\dots,m+1;a,b\rangle_{m+1;a,b}}\nonumber\\
&=&\sum_{k\ne i,j}\mathcal{D}_{ij,k}(p_1,\dots,p_{m+1};p_a,p_b)\nonumber\\
&+&\left[\mathcal{D}_{ij}^a(p_1,\dots,p_{m+1};p_a,p_b)+\mathcal{D}_{ij}^b(p_1,\dots,p_{m+1};p_a,p_b)\right],
\label{init}
\end{eqnarray}
where we have neglected all finite terms. The first term on the RHS is a dipole contribution with a final-state emitter and a final-state spectator and is given by
\begin{eqnarray}
\lefteqn{\mathcal{D}_{ij,k}(p_1,\dots,p_{m+1};p_a,p_b)=-\frac{1}{2p_i.p_j}}\nonumber\\
&\times_{m;a,b}\langle 1,\dots,\widetilde{ij},\dots,\widetilde{k},\dots,m+1;a,b|\frac{{\bf T}_{ij}.{\bf T}_k}{{\bf T}_{ij}^2}{\bf V}_{ij,k}|1,\dots,\widetilde{ij},\dots,\widetilde{k},\dots,m+1;a,b\rangle_{m;a,b}.
\phantom{(9)}
\label{FF}
\end{eqnarray}
The $m$-parton matrix element on the RHS of Eq.~(\ref{FF}) is obtained from the $m+1$-parton matrix element by replacing (a) the partons $i$ and $j$ with a single parton $\widetilde{ij}$ (the emitter) and (b) the parton $k$ with the parton $\widetilde{k}$ (the spectator).

The quantum numbers (except momenta) of the partons of the $m$-parton matrix element are assigned as follows. The quantum numbers of all partons except the emitter and the spectator are as in the $m+1$-parton matrix element. The quantum numbers of the spectator, $\widetilde{k}$ are the same as $k$. The quantum numbers of the emitter, $\widetilde{ij}$, are obtained  according to their conservation in the collinear splitting process $\widetilde{ij}\rightarrow i+j$, i.e.\ if $i$ and $j$ are something + gluon then $\widetilde{ij}$ is something; if $i$ and $j$ are quark + antiquark then $\widetilde{ij}$ is a gluon. 

The momenta of the emitter and spectator are defined as follows
\begin{eqnarray}
\widetilde{p}^{\mu}_k=\frac{1}{1-y_{ij,k}}p^{\mu}_k,\qquad\widetilde{p}^{\mu}_{ij}=p^{\mu}_i+p^{\mu}_j-\frac{y_{ij,k}}{1-y_{ij,k}}p^{\mu}_k,
\end{eqnarray}
where the scalar $y_{ij,k}$ is given by
\begin{equation}
y_{ij,k}=\frac{p_i.p_j}{p_i.p_j+p_j.p_k+p_k.p_i}.
\end{equation}
Note that the emitter and spectator are both on mass-shell and in the splitting process $\{\widetilde{ij},\widetilde{k}\}\rightarrow\{i,j,k\}$ momentum conservation is implemented exactly.

Note that the matrix element squared does not factorize exactly on the RHS of Eq.~(\ref{FF}), there are colour and spin correlations. ${\bf T}_{ij}$ and ${\bf T}_k$ are the colour charges of the emitter and spectator respectively, and ${\bf V}_{ij,k}$ is an operator in the helicity space of the emitter, which is related to the $d$-dimensional Altarelli-Parisi splitting functions. As we wish only to outline the general method rather than go into every detail we shall not list here the splitting matrices for each type of dipole contribution. We shall write out these matrices for the case of final-state emitter and spectator, but for the other three types we refer the reader to~\cite{Catani:1996vz}.

These splitting functions are given in terms of the Lorentz scalars $y_{ij,k}$, $\widetilde{z}_i$ and $\widetilde{z}_j$ given by
\begin{eqnarray}
\widetilde{z}_i&=&\frac{p_i.p_k}{p_j.p_k+p_i.p_k}=\frac{p_i.\widetilde{p}_k}{\widetilde{p}_{ij}.\widetilde{p}_k},\nonumber\\
\widetilde{z}_j&=&\frac{p_j.p_k}{p_j.p_k+p_i.p_k}=\frac{p_j.\widetilde{p}_k}{\widetilde{p}_{ij}.\widetilde{p}_k}=1-\widetilde{z}_i.
\end{eqnarray}
For a fermion + gluon splitting we have
\begin{eqnarray}
\langle s|{\bf V}_{q_ig_j,k}(\widetilde{z}_i;y_{ij,k})|s^{\prime}\rangle &=& 8\pi\mu^{2\epsilon}\alpha_SC_F\left[\frac{2}{1-\widetilde{z}_i(1-y_{ij,k})}-(1+\widetilde{z}_i)-\epsilon(1-\widetilde{z}_i)\right]\delta_{ss^{\prime}}\nonumber\\
&=&V_{q_ig_j,k}\delta_{ss^{\prime}},
\end{eqnarray}
where $s$ and $s^{\prime}$ are the spin indices of the fermion $\widetilde{ij}$ in the vector $|\dots,\widetilde{ij},\dots\rangle$. For quark + antiquark and gluon + gluon splitting we have
\begin{eqnarray}
\langle\mu|{\bf V}_{q_i\bar{q}_j,k}(\widetilde{z}_i)|\nu\rangle &=& 8\pi\mu^{2\epsilon}\alpha_ST_R\left[-g^{\mu\nu}-\frac{2}{p_i.p_j}(\widetilde{z}_ip_i^{\mu}-\widetilde{z}_jp_j^{\mu})(\widetilde{z}_ip_i^{\nu}-\widetilde{z}_jp_j^{\nu})\right]\nonumber\\
&=&V_{q_i\bar{q}_j,k}^{\mu\nu},
\end{eqnarray}
\begin{eqnarray}
\langle\mu|{\bf V}_{g_ig_j,k}(\widetilde{z}_i;y_{ij,k})|\nu\rangle &=& 16\pi\mu^{2\epsilon}\alpha_SC_A\left[-g^{\mu\nu}\left(\frac{1}{1-\widetilde{z}_i(1-y_{ij,k})}+\frac{1}{1-\widetilde{z}_j(1-y_{ij,k})}-2\right)\right.\nonumber\\
&&\left.+(1-\epsilon)\frac{1}{p_i.p_j}(\widetilde{z}_ip_i^{\mu}-\widetilde{z}_jp_j^{\mu})(\widetilde{z}_ip_i^{\nu}-\widetilde{z}_jp_j^{\nu})\right]\nonumber\\
&=&V_{g_ig_j,k}^{\mu\nu},
\end{eqnarray}
 where $\mu$ and $\nu$ are the spin indices of the gluon $\widetilde{ij}$ in the vector $|\dots,\widetilde{ij},\dots\rangle$. 

The other two terms on the RHS of Eq.~(\ref{init}) are dipole contributions with a final-state emitter and an initial-state spectator. This is given by
\begin{eqnarray}
\lefteqn{\mathcal{D}_{ij}^a(p_1,\dots,p_{m+1};p_a,p_b)=-\frac{1}{2p_i.p_j}\frac{1}{x_{ij,a}}}\nonumber\\
&\times_{m;a,b}\langle 1,\dots,\widetilde{ij},\dots,m+1;\widetilde{a},b|\frac{{\bf T}_{ij}.{\bf T}_a}{{\bf T}_{ij}^2}{\bf V}_{ij}^a|1,\dots,\widetilde{ij},\dots,m+1;\widetilde{a},b\rangle_{m;a,b},
\label{FI}
\end{eqnarray}
where the Lorentz scalar $x_{ij,a}$ is given by
\begin{equation}
x_{ij,a}=\frac{p_i.p_a+p_j.p_a-p_i.p_j}{(p_i+p_j).p_a}.
\end{equation}

All quantum numbers except the momenta of the emitter and spectator are given as in the case of the dipole term $\mathcal{D}_{ij,k}$. These are given by
\begin{equation}
\widetilde{p}^{\mu}_a=x_{ij,a}p_a^{\mu},\qquad\widetilde{p}^{\mu}_{ij}=p^{\mu}_{i}+p^{\mu}_{j}-(1-x_{ij,a})p^{\mu}_{a}.
\end{equation}
For a definition of the splitting matrices ${\bf V}_{ij}^a$ we refer the reader to the original paper\cite{Catani:1996vz}.
\subsubsection{Initial-state singularities}
\label{initial}
Using the dipole factorization formula we can write the real matrix element squared as we approach an initial-state singularity ($p_a.p_i\rightarrow 0$) as
\begin{eqnarray}
\lefteqn{_{m+1;a,b}\langle1,\dots,m+1;a,b|1,\dots,m+1;a,b\rangle_{m+1;a,b}}\nonumber\\
&=&\sum_{k\ne i}\mathcal{D}_{k}^{ai}(p_1,\dots,p_{m+1};p_a,p_b)+\mathcal{D}^{ai,b}(p_1,\dots,p_{m+1};p_a,p_b).
\label{final}
\end{eqnarray}
where again we neglect all finite terms. The first new dipole term is given by 
\begin{eqnarray}
\lefteqn{\mathcal{D}_{k}^{ai}(p_1,\dots,p_{m+1};p_a,p_b)=-\frac{1}{2p_a.p_i}\frac{1}{x_{ik,a}}}\nonumber\\
&\times_{m;a,b}\langle 1,\dots,\widetilde{k},\dots,m+1;\widetilde{ai},b|\frac{{\bf T}_{ai}.{\bf T}_k}{{\bf T}_{ai}^2}{\bf V}_{ai}^k|1,\dots,\widetilde{k},\dots,m+1;\widetilde{ai},b\rangle_{m;a,b},
\label{IF}
\end{eqnarray}
where the Lorentz scalar $x_{ik,a}$ is given by
\begin{equation}
x_{ik,a}=\frac{p_k.p_a+p_i.p_a-p_i.p_k}{(p_k+p_i).p_a},
\end{equation}
and the momenta of the emitter and spectator are given by
\begin{equation}
\widetilde{p}^{\mu}_{ai}=x_{ik,a}p^{\mu}_a,\qquad \widetilde{p}^{\mu}_{k}=p^{\mu}_k+p^{\mu}_i-(1-x_{ik,a})p^{\mu}_a.
\end{equation}
The second type of dipole on the RHS of Eq.~(\ref{final}) is given by
\begin{eqnarray}
\lefteqn{\mathcal{D}^{ai,b}(p_1,\dots,p_{m+1};p_a,p_b)=-\frac{1}{2p_a.p_i}\frac{1}{x_{i,ab}}}\nonumber\\
&\times_{m;a,b}\langle \widetilde{1},\dots,\widetilde{m+1};\widetilde{ai},b|\frac{{\bf T}_{b}.{\bf T}_{ai}}{{\bf T}_{ai}^2}{\bf V}^{ai,b}|\widetilde{1},\dots,\widetilde{m+1};\widetilde{ai},b\rangle_{m;a,b},
\label{II}
\end{eqnarray}
where the Lorentz scalar $x_{i,ab}$ is given by
\begin{equation}
x_{i,ab}=\frac{p_a.p_b-p_i.p_a-p_i.p_b}{p_a.p_b}.
\end{equation}
The structure of this dipole term differs from the others since it is convenient to leave the momentum of the spectator, $p_b$, unchanged. Thus the tilde momenta phase point at which the matrix element $|\widetilde{1},\dots,\widetilde{m+1};\widetilde{ai},b\rangle_{m;a,b}$ is evaluated is found as follows. The momentum of the emitter is parallel to $p_a$:
\begin{equation}
\widetilde{p}_{ai}^{\mu}=x_{i,ab}p_a^{\mu},
\end{equation}
and in order to implement momentum conservation exactly all other final-state momenta must recoil as follows:
\begin{equation}
\widetilde{k}^{\mu}_j=k^{\mu}_j-\frac{2k_j.(K+\widetilde{K})}{(K+\widetilde{K})^2}(K+\widetilde{K})^{\mu}+\frac{2k_j.K}{K^2}\widetilde{K}^{\mu},
\end{equation}
where
\begin{eqnarray}
K^{\mu}&=&p_a^{\mu}+p_b^{\mu}-p_i^{\mu},\nonumber\\
\widetilde{K}^{\mu}&=&\widetilde{p}_{ai}^{\mu}+p_b^{\mu}.
\end{eqnarray}
Note that it is all final-state particles that recoil, not just the QCD partons. Like all the previous dipole contributions, all particles remain on mass-shell and momentum conservation is exact.
\subsubsection{Summary}
\label{summ}
We have stated that each of these dipole terms acts as a counter term for the real matrix element squared, $|\mathcal{M}_{m+1;a,b}|^2$, in a different singular region of phase space. The sum of all dipole terms will therefore act as a local counter term in all singular regions of phase space. We need to construct a counter term for $d\sigma^R$, where
\begin{equation}
d\sigma^R\propto d\phi^{(m+1)}(p;Q)|\mathcal{M}_{m+1;a,b}|^2(p)F_J^{(m+1)}(p),
\end{equation}
and it is clear that such a counter term, $d\sigma^A$, must contain a factor of the jet function that defines the observable, $F_J$.

Recall that the cancellation of divergences between the real and virtual contribution was only guaranteed for infrared safe observables defined by the following property of the jet function:
\begin{equation}
F_J^{(m+1)}\rightarrow F_J^{(m)}
\end{equation}
as one approaches an $m+1$-parton configuration that is kinematically degenerate with an $m$-parton configuration. Since $d\sigma^A$ must match the singular behaviour of $d\sigma^R$ and in the singular regions of phase space we know that $F_J^{(m+1)}=F_J^{(m)}$, it appears we have a choice when constructing the counter-term $d\sigma^A$: it can be constructed by multiplying each of the dipole terms by either $F_J^{(m+1)}(p)$ or $F_J^{(m)}(\widetilde{p})$. Either would give a counter term that matches the singularities of the real cross section, but only the latter gives a finite cross section that can be implemented by Monte Carlo methods, as we shall now discuss.

An obvious requirement of the counter term is that it matches the singular behaviour of the real term, but does not contain any additional singularities. The dipole terms are all proportional to the square of the Born level matrix element\footnote{Actually it is the relevant colour and spin projection of the Born level matrix element squared, but that does not affect this argument.} evaluated at the $m$-parton phase space point given by the tilde momenta $\{\widetilde{p}\}$. Multiplying this is a factor that diverges in the soft and collinear regions of phase space ($1/p_i.p_j$). Consider some point in $m+1$-parton phase space for which the real matrix element is finite and $F_J^{(m+1)}(p)$ is non-zero; it is possible that one of the phase space mappings onto $m$-parton phase space produces an $m$-parton configuration arbitrarily close to a singular region of phase space such that $|\mathcal{M}^B_{m;a,b}|^2$ is arbitrarily large. For such a point in phase space $d\sigma^R$ will be finite and the first choice that $d\sigma^A$ is proportional to $F_J^{(m+1)}(p)$ will be arbitrarily large, i.e.\ $d\sigma^A$ will contain extra (`spurious') singularities.

However recall that, by definition, the Born level contribution to any observable given by Eq.~(\ref{born}) is finite. Thus the jet function $F_J^{(m)}$ must be defined such that as the Born level matrix element diverges $F_J^{(m)}\rightarrow 0$. Therefore if we make the second choice that $d\sigma^A$ is proportional to $F_J^{(m)}(\widetilde{p})$ these extra singularities will be removed.

Finally we define the counter term as
\begin{eqnarray}
d\sigma^A=d\phi^{(m+1)}(p;Q)\sum_{{\rm dipoles}}(\mathcal{D}\cdot F^{(m)})(p)
\end{eqnarray}
where
\begin{eqnarray}
\sum_{{\rm dipoles}}(\mathcal{D}\cdot F^{(m)})(p)&=&\left\{\sum_{i,j}\left[\sum_{k\ne i,j}\mathcal{D}_{ij,k}(p)F^{(m)}_J(\widetilde{p})+\mathcal{D}^a_{ij}(p)F^{(m)}_J(\widetilde{p})+\mathcal{D}^b_{ij}(p)F^{(m)}_J(\widetilde{p})\right]\right.\nonumber\\
&&\left.+\sum_i\left[\sum_{k\ne i}\mathcal{D}^{ai}_{k}(p)F^{(m)}_J(\widetilde{p})+\mathcal{D}^{ai,b}(p)F^{(m)}_J(\widetilde{p})+(a\leftrightarrow b)\right]\right\}.
\end{eqnarray}

\subsection{Integrated dipole terms}
\label{end}
In this section we shall discuss the terms that result from the integration of the dipole terms over the single parton subspace leading to the singularities. We shall not give a detailed discussion, but will concentrate on those aspects that will be useful for the user to include a new process.

In section~\ref{hadron} we outlined the basic idea that some of the divergence from dipole terms with initial state partons cancels the $1/\epsilon$ poles of the collinear subtraction term to give the {\it finite} convolution term in Eq.~(\ref{conv}). The rest of the divergence is contained in the insertion operator and we stated that this will cancel the divergence of the virtual contribution such that the integral
\begin{eqnarray}
\sigma^{{\rm NLO}\{m\}}(p)=\int_m\left[d\sigma^V(p)+d\sigma^B(p)\otimes{\bf I}\right]_{\epsilon=0},
\end{eqnarray}
is finite. After the integration over the single particle subspace no spin correlations survive, so the insertion operator is an operator only in colour space. The second term on the RHS is
\begin{equation}
d\sigma^B(p)\otimes{\bf I}=_{m,a,b}\langle 1,\dots,m;a,b|{\bf I}(\{p\};\epsilon)|1,\dots,m;a,b\rangle_{m,a,b},
\end{equation}
where the insertion operator is given by
\begin{eqnarray}
{\bf I}(\{p\};\epsilon)&=&-\frac{\alpha_S}{2\pi}\frac{1}{\Gamma(1-\epsilon)}\sum_{I}\frac{1}{{\bf T}^2_I}\mathcal{V}_I(\epsilon)\sum_{J\ne I}{\bf T}_I\cdot{\bf T}_J\left(\frac{4\pi\mu^2}{2p_I\cdot p_J}\right)^{\epsilon}.
\end{eqnarray}
The sums runs over both initial-state and final-state partons, and the flavour-dependent universal singular functions $\mathcal{V}_I(\epsilon)$ are given by
\begin{eqnarray}
\mathcal{V}_q(\epsilon)&=&C_F\left[\frac{1}{\epsilon^2}+\frac{2}{3\epsilon}+5-\frac{\pi^2}{2}+{\cal O}(\epsilon)\right],\nonumber\\
\mathcal{V}_g(\epsilon)&=&\frac{C_A}{\epsilon^2}+\left(\frac{11}{6}C_A-\frac{2}{3}T_RN_f\right)\frac{1}{\epsilon}+C_A\left(\frac{50}{9}-\frac{\pi^2}{2}\right)-T_RN_f\frac{16}{9}+{\cal O}(\epsilon).
\end{eqnarray}
Since the singularities of this insertion operator term cancel with those of the renormalized one-loop matrix element, \texttt{TeVJet} needs only to calculate the {\it finite} parts of each. What we mean by the finite parts warrants a little explanation. The insertion operator can be re-written as
\begin{eqnarray}
{\bf I}(\{p\};\epsilon)&=&-\frac{\alpha_S}{2\pi}\frac{1}{\Gamma(1-\epsilon)}\left(\frac{4\pi\mu^2}{Q^2}\right)^{\epsilon}\sum_{I}\sum_{J\ne I}\frac{{\bf T}_I\cdot{\bf T}_J}{{\bf T}^2_I}\mathcal{V}_I(\epsilon)\left(\frac{Q^2}{2p_I\cdot p_J}\right)^{\epsilon},
\end{eqnarray}
where $Q$ is some arbitrary scale. Using dimensional regularization the renormalized one-loop matrix element in the $\overline{\mathrm{MS}}$ scheme is typically of the form
\begin{equation}
|\mathcal{M}|^2_{1-loop}=\frac{\alpha_S}{2\pi}\frac{1}{\Gamma(1-\epsilon)}\left(\frac{4\pi\mu^2}{Q^2}\right)^{\epsilon}\left\{\frac{A}{\epsilon^2}+\frac{B}{\epsilon}+C+{\cal O}(\epsilon)\right\},
\label{1loop}
\end{equation}
where $\mu$ is the dimensional regularization scale (which, in common with Ref.~\cite{Catani:1996vz}, we set equal to the renormalization scale throughout), and $Q$ is some scale relevant to the process~-- the precise definition of $Q$ fixes the precise form of the functions $B$ and $C$. The convention in \texttt{TeVJet} is to extract a factor of
\begin{equation}
\frac{1}{\Gamma(1-\epsilon)}\left(\frac{4\pi\mu^2}{Q^2}\right)^{\epsilon},
\end{equation}
from the insertion operator and the one-loop matrix element, and to calculate the finite part of what is left. The reference scale $Q$ can be chosen by the user. It can be any scale as long as the user is consistent, and the same scale is extracted when calculating the one-loop matrix element. Thus for the insertion operator counter-term \texttt{TeVJet} needs to calculate the finite part of terms of the form
\begin{equation}
\mathcal{V}_I(\epsilon)\left(\frac{Q^2}{2p_I\cdot p_J}\right)^{\epsilon},
\label{ins}
\end{equation}
where the singular functions $\mathcal{V}_I(\epsilon)$ have the form
\begin{equation}
\mathcal{V}_I(\epsilon)=\frac{D}{\epsilon^2}+\frac{E}{\epsilon}+F+{\cal O}(\epsilon),
\end{equation}
and the second term of Eq.~(\ref{ins}) can be expanded as a Taylor series
\begin{equation}
a^{\epsilon}=1+\frac{\log(a)}{1!}\epsilon+\frac{\log^2(a)}{2!}\epsilon^2+\dots.
\end{equation}
Thus the finite part of Eq.~(\ref{ins}) is given by:
\begin{equation}
F+\log\left(\frac{Q^2}{2p_I\cdot p_J}\right)\frac{E}{1!}+\log^2\left(\frac{Q^2}{2p_I\cdot p_J}\right)\frac{D}{2!}.
\end{equation}
In order to put a new process into \texttt{TeVJet} the user must provide code that fixes $Q$ and returns the finite part of Eq.~(\ref{1loop}) after the common factor has been removed, i.e.
\begin{equation}
\frac{\alpha_S}{2\pi}C.
\end{equation}

\section{\texttt{TeVJet} Design}
\label{des}

\texttt{TeVJet} is a framework for calculating jet cross sections in NLO QCD using the dipole subtraction method~\cite{Catani:1996vz}. It has been designed to make the inclusion of a new scattering process as straightforward as possible. The user must supply the usual ingredients for an NLO calculation: the Born level and real emission matrix elements, and the interference between the Born level and one-loop amplitudes. From these \texttt{TeVJet} automatically calculates the subtraction terms and various insertion operators; and performs the integrations over phase space.

\texttt{TeVJet} has been written in a modular way using C++; the library is made of the following classes\footnote{Note: all of these classes are declared inside \texttt{namespace TeVJet}.}, and classes that inherit from them. The main classes are:
\begin{itemize}
\item \texttt{Calculation}
\item \texttt{Process}
\item \texttt{JetFunction}
\item \texttt{Tools}
\item \texttt{PhaseSpace}
\end{itemize}
The following classes are available to the user for producing histograms:
\begin{itemize}
\item \texttt{Histogram}
\item \texttt{HistoFile}
\end{itemize}
The following classes contain the process-dependent parts of the calculation:
\begin{itemize}
\item \texttt{Amplitude}
\item \texttt{MatrixElementSquared}
\item \texttt{LoopIntegral}
\end{itemize}
The following classes contain the process-independent parts of the calculation:
\begin{itemize}
\item \texttt{SubtractionTerm}
\item \texttt{InsertionOperator}
\item \texttt{SplittingKernelP}
\item \texttt{SplittingKernelK}
\end{itemize}

\begin{figure}[t]
\includegraphics[width=\textwidth]{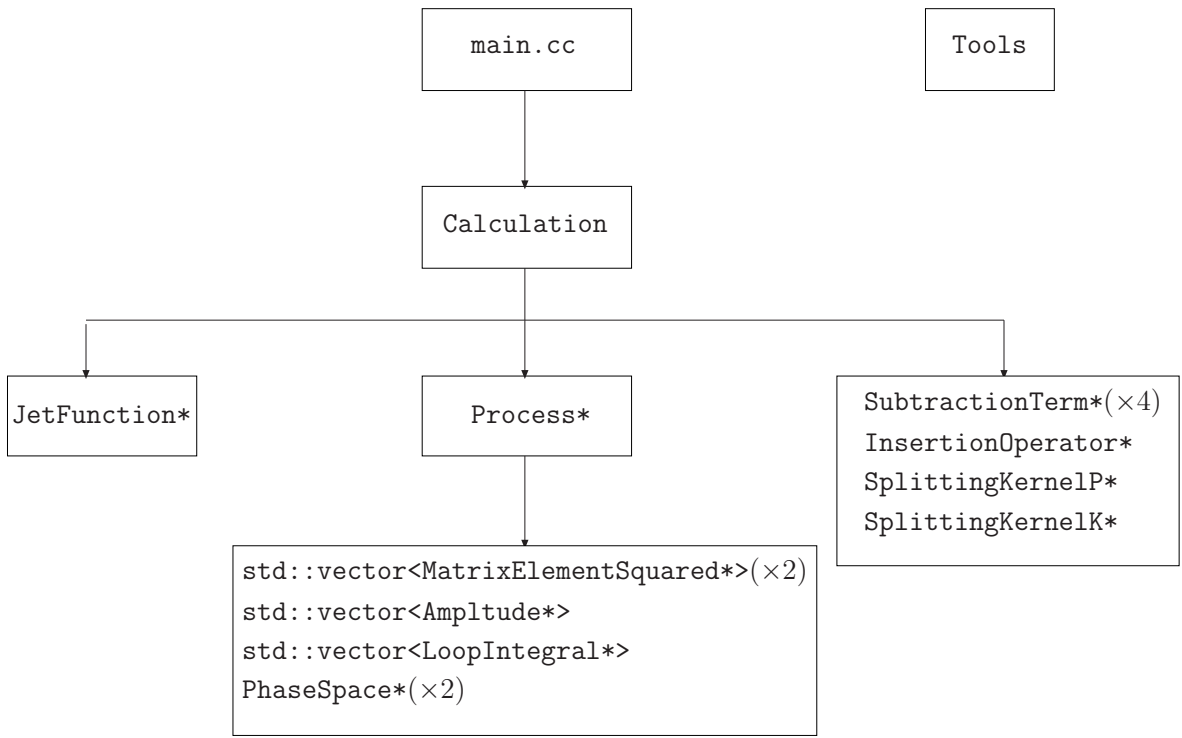}
\caption{Schematic diagram showing the structure of \texttt{TeVJet}.}
\label{diagram}
\end{figure}

In section~\ref{main.cc} we give an example of a main program. The main driving class is the \texttt{Calculation} class, which takes, on construction, a pointer to a \texttt{Process} and a pointer to a \texttt{JetFunction}. The \texttt{Process} class contains all of the process dependent parts of the calculation, whereas the \texttt{JetFunction} contains the definition of the observable(s) to be calculated. Figure~\ref{diagram} shows schematically the structure of the program.
In this section we shall provide a description of some of these classes, with particular emphasis on the methods that are needed by the user. In sections~\ref{calc}--\ref{phase} we describe the main five classes listed above. In sections~\ref{histo1}--\ref{histo2} we describe the classes used to create histograms. In sections~\ref{mat}--\ref{loop} we describe the classes that store all of the process-dependent parts of the calculation. Some knowledge of these is required if the user wishes to include a new process in \texttt{TeVJet}. The user should not need to know the details of the four classes that contain the process-independent parts of the calculation, and they are not described here.

At this point we distinguish two types of user:
\begin{itemize}
\item A user who wishes to do a new calculation by writing a new \texttt{Process} class. They will need some knowledge of all of the \texttt{TeVJet} classes.
\item A user who wishes to use \texttt{TeVJet} to calculate rates, plot distributions and so on for an existing process. In this case they should only need detailed knowledge of the \texttt{JetFunction} class and the two histogram classes, \texttt{Histogram} and \texttt{HistoFile}, with some knowledge of the other main classes.
\end{itemize}

\subsection{The Calculation class}
This is the driving class of the calculation. It collects together all of the parts of the calculation and performs the phase space integrals and convolutions with the parton distribution functions. On construction it is passed all of the process-dependent parts of the calculation. It also has, as private members, pointers to all of the process-independent parts, i.e.\ the four \texttt{SubtractionTerm}s, and the \texttt{InsertionOperator}, \texttt{SplittingKernelP} and \texttt{SplittingKernelK}.

\subsubsection*{Calculation methods}
\label{calc}
\begin{verbatim}
    Calculation(TeVJet::Process * proc, TeVJet::JetFunction * jet);
\end{verbatim}
This is the constructor for the \texttt{Calculation} class. 

\begin{verbatim}
    void Initialize();
\end{verbatim}
This initializes the beam momenta, the random number generator, the \texttt{Process} and, if there are no hadrons in the initial state, the pointers to the two \texttt{PhaseSpace} objects.

The following set functions should be called to set the beam energies in units of GeV and the random number seed before \texttt{initialize} otherwise default parameters will be used. The default beam energies depend on the number of initial state hadrons in the \texttt{Process}~-- for zero hadrons the beam energies are both set to $M_Z/2$; for one hadron the hadron energy is set to 920.0~GeV and the lepton energy to 27.5~GeV; for two hadrons the energies are both set to 7000.0~GeV.
\begin{verbatim}
    void setEbeam1(double Ebeam1);
    void setEbeam2(double Ebeam1);
    void setSeed(long seed);
\end{verbatim}
The integral of the Born level result of Eq.~(\ref{bornloop}) can be evaluated by calling the method
\begin{verbatim}
    void BornLoop();
\end{verbatim}
and the three integrals of Eq.~(\ref{nloloop}) can be evaluated by calling the methods
\begin{verbatim}
    void RealLoop();
    void VirtLoop();
    void ConvLoop();
\end{verbatim}
The following additional set functions are available to set the number of phase space points to choose for these four integrals.
\begin{verbatim}
    void setNborn(int Nborn);
    void setNreal(int Nreal);
    void setNvirt(int Nvirt);
    void setNconv(int Nconv);
\end{verbatim}
When a new process is written (see section~\ref{proc}) the user must choose a renormalization and factorization scale. This must be an infrared safe function of the momenta, e.g.\ the total CM energy in the case of $e^+e^-$ collisions. The \texttt{Calculation} will multiply each of these scales by a number that can be set by the user via the following methods:
\begin{verbatim}
    void setRenScaleFactor(double factor);
    void setFacScaleFactor(double factor);
\end{verbatim}
Each of these numbers is set to 1.0 by default.
\begin{verbatim}
    void setRenScheme(std::string scheme);
    void setFacScheme(std::string scheme);
\end{verbatim}
These methods are used to set the renormalization and factorization schemes. At present only one renormalization scheme is available, and the only possible value of scheme is \texttt{scheme=}``\texttt{MSbar}''. Two factorization schemes are available by setting \texttt{scheme=}``\texttt{MSbar}'' or \texttt{scheme=}``\texttt{DIS}''. It is important to ensure that the same factorization scheme has been used to extract the pdf, i.e.\ the user must choose a pdf that uses either the $\overline{\mathrm{MS}}$ or the DIS factorization scheme, and then should set \texttt{TeVJet} to use the same scheme.
\begin{verbatim}
    void InitPDF(std::string pdfname, int pdfmemset);
\end{verbatim}
\texttt{TeVJet} links to LHApdf for the parton distribution functions. The user may use this method to choose the pdf set, and the member of that set that they wish to use. If this is not called the default set (cteq61) is used.

\subsection{The Process class}
\label{proc}
The purpose of the \texttt{Process} class is to pass all of the process-dependent parts (e.g.\ matrix elements) to the \texttt{Calculation} class. It is an abstract base class, which has a number of virtual methods. All of the process-dependent parts of the calculation are passed to the \texttt{Calculation} class via a class that inherits from the \texttt{Process} class and in which these methods have been over-written. The process-dependent parts of the calculation are stored in the public members:
\begin{verbatim}
    std::vector<MatrixElementSquared*> MatrixBorn;
    std::vector<MatrixElementSquared*> MatrixReal;
    std::vector<Amplitude*> BornAmp;
    std::vector<LoopIntegral*> Loops;
\end{verbatim}
We define these classes later, but for now make one comment about them: A single process can contain several subprocesses, implemented by different members of these vectors. For example the process $e^+e^-\rightarrow 3$ jets will contain subprocesses such as $e^+e^-\rightarrow u\bar{u}g$.

The process class also has two pointers to \texttt{PhaseSpace} objects:
\begin{verbatim}
    PhaseSpace * PhaseBorn;
    PhaseSpace * PhaseReal;
\end{verbatim}
\subsubsection*{Process methods}
When writing a class that inherits from the \texttt{Process} class, the following virtual methods must be written.
\begin{verbatim}
    virtual void   Initialize()=0;
\end{verbatim}
This is called by the \texttt{Calculation} method \texttt{Initialize()}. The user should call the constructors of the two \texttt{PhaseSpace} classes. The user should also call the constructors of all of the Born level \texttt{Amplitude} classes, the Born and real \texttt{MatrixElementSquared} classes and the \texttt{LoopIntegral} classes. These can either be existing classes, or the user can write their own (see relevant sections below).
\begin{verbatim}
    virtual int    NhadronBeams()=0;
\end{verbatim}
This should be set to 0 for processes with no hadrons in the initial state, 1 for DIS processes, and 2 for hadron-hadron processes.
\begin{verbatim}
    virtual bool   DoNLO()=0;
\end{verbatim}
This returns a flag that may be set to \texttt{false} if the user only wishes to calculate the Born level result for a process. 
\begin{verbatim}
    virtual double getRenScale(
                    std::vector<HepLorentzVector> Momenta)=0;
    virtual double getFacScale(
                    std::vector<HepLorentzVector> Momenta)=0;
\end{verbatim}
These methods are called by the \texttt{Calculation} class to provide the renormalization and factorization scales respectively and the user should supply code to return an infrared safe function of the input momenta. This could be for example the total CM energy in $e^+e^-$ collisions, or the virtuality of the exchanged vector boson in DIS. The first two elements of the input vector are the momenta of the initial state partons, followed by the final state partons.
\begin{verbatim}
    virtual double getRefScale(
                    std::vector<HepLorentzVector> Momenta)=0;
\end{verbatim}
As explained in section~\ref{end}, the convention in \texttt{TeVJet} for defining the finite parts of the insertion operator and the one-loop matrix element is to extract a common factor of
\begin{equation}
\frac{1}{\Gamma(1-\epsilon)}\left(\frac{4\pi\mu^2}{Q^2}\right)^{\epsilon},
\end{equation}
and drop the divergent terms in $\epsilon$. The reference scale $Q$ can be chosen by the user and this is what the method \texttt{getRefScale} should return. It can be any scale as long as the user is consistent and the same scale is extracted when calculating the one-loop matrix element.
 
\subsection{The JetFunction class}
\label{jet}
The purpose of the \texttt{JetFunction} class is to calculate the value of the jet function $F_J$ introduced in Eq.~(\ref{born}). It is an abstract base class, which has a number of virtual methods. The main method is the user defined function \texttt{PlotStuff}, which is called several times during the run. \texttt{TeVJet} passes each phase space point with a weight to the user, and the user must provide the code to calculate the observable(s) of interest. For example to calculate a differential cross section, $d\sigma/dX$, one should calculate $X$ from the final state momenta and fill a histogram of $X$ with the weight.

The method \texttt{PlotStuff} is called several times per `event' in \texttt{RealLoop}: once for each $m+1$-body phase space point, and once for each of the different $m$-body phase space points needed for each subtraction term (see discussion in section~\ref{summ}). In general the error on an observable (e.g.\ a single bin of a histogram) can be obtained from a weighted distribution of events by adding their weights in quadrature. However in the subtraction algorithm there can be large cancellations between the real matrix element and the subtraction terms and, since these are correlated, one should sum them first before using them to calculate the error on an observable. Therefore if one is using a histogramming package that automatically calculates the errors on each bin using the squares of the weights one will overestimate the errors if the real matrix element and each subtraction term are filled separately. In order to avoid this there is a method \texttt{Terminate}, which is called at the end of each full event. The user should fill a temporary histogram in \texttt{PlotStuff} each bin of which is added to the main histogram in \texttt{Terminate}. The same applies to the convolution term calculated by \texttt{ConvLoop}~-- \texttt{PlotStuff} is called twice with two correlated phase space points followed by one call to \texttt{Terminate}.

Since the details of the observable(s) to be calculated are contained in a class that inherits from the \texttt{JetFunction} class and in which these methods have been over-written, we describe the methods below.

\subsubsection*{JetFunction methods}
The constructor of the \texttt{JetFunction} class takes a \texttt{std::string}, which is the name of the output file:
\begin{verbatim}
    JetFunction(std::string filename);
\end{verbatim}
The \texttt{JetFunction} class has a standard vector of pointers to \texttt{Histogram} objects:
\begin{verbatim}
    std::vector<Histogram*> m_histovec;
\end{verbatim}
and a pointer to a \texttt{HistoFile} object:
\begin{verbatim}
    HistoFile * m_outfile; 
\end{verbatim}
The following method will write all of the histograms stored in \texttt{m\_histovec} to the output file:
\begin{verbatim}
    void WriteFile();
\end{verbatim}
When writing a class that inherits from the \texttt{JetFunction} class, the following virtual methods must be written.

\begin{verbatim}
    virtual bool KeepPhasePoint(
                     std::vector<HepLorentzVector> finals)=0;
\end{verbatim}
This method is called after the generation of each phase space point, before the weight is calculated. If this returns \texttt{true} the point is kept and the weight calculated, if it returns \texttt{false} the point is rejected and the weight never calculated. If the user wishes to put some cuts on phase space they should be implemented here.
\begin{verbatim}
    virtual void PlotStuff(double weight, 
                           std::vector<HepLorentzVector> finals,
                           std::string type)=0;
\end{verbatim}
This method is called several times by the \texttt{Calculation} class. The variable \texttt{type} indicates which type of contribution is being calculated. It is called once for each phase space point in \texttt{BornLoop} and \texttt{VirtLoop} (\texttt{type} is \texttt{born} or \texttt{virtual} respectively); several times for each phase space point in \texttt{RealLoop}~-- once for the real emission $|\mathcal{M}|^2$ (\texttt{type} is \texttt{real}) and once for each dipole configuration (\texttt{type} is \texttt{subtraction}); and twice for each phase space point in \texttt{ConvLoop} (\texttt{type} is \texttt{convolution}). Here the observable(s) should be calculated using the final state momenta and the weight. For example to calculate a differential cross section, $d\sigma/dX$, one should calculate $X$ from the final state momenta and fill a histogram of $X$ with the \texttt{weight}.
\begin{verbatim}
    virtual void Terminate()=0;
\end{verbatim}
As explained above, the phase space points for the real emission $|\mathcal{M}|^2$ and the dipole configurations are correlated and one will over-estimate the error if one fills the histogram once for this `event' and once for each `counter-event'. One way to avoid this is to fill a dummy histogram in \texttt{PlotStuff} in this way, and at the end of the `full event' to copy this histogram into the main histogram. The method \texttt{Terminate()} gets called once per real phase space point in \texttt{RealLoop} for this purpose. It is also called once per full `event' in \texttt{ConvLoop}.

For this purpose the following method exists to copy the contents of one histogram (\texttt{histo2}) into another histogram (\texttt{histo1}):
\begin{verbatim}
    void FillEvent(Histogram * histo1, Histogram * histo2);
\end{verbatim}

\subsection{The Tools class}
\label{tools}
All parts of the program need access to the Standard Model parameters (for which, we note, we use natural units based on GeV, with $c=1$).  We therefore provide them through the \texttt{Tools} class, which is a `singleton' class, as we now describe.

\subsubsection*{The singleton pattern}
The `singleton pattern' is essentially the object oriented way of declaring a global variable. A singleton class has a {\it private} constructor, but also has as a static private member a pointer to itself that can be accessed by a public method. Here is a skeleton of the class declaration:
\begin{verbatim}
  class Tools{
  public:
    ~Tools();    
    static Tools * getInstance();

  private:
    //private constructor
    Tools();

    static bool instance;
    static Tools * g_tools;

  };// end class Tools
\end{verbatim}
The method \texttt{getInstance} returns the private pointer (on the first call the constructor is called):
\begin{verbatim}
TeVJet::Tools * TeVJet::Tools::getInstance(){
  if(!instance){
    g_tools = new TeVJet::Tools();
    instance = true;
    return g_tools;
  }
  else{
    return g_tools;
  }
}
\end{verbatim}
The \texttt{Tools} class is therefore a safe global variable that stores the values of the SM parameters. It does not get created until one needs it for the first time, at which point the constructor is called and the default values for the parameters are set.
\subsubsection*{Tools methods}
Through the methods of the \texttt{Tools} class the user may access the parameters, or set them to values other than the defaults. 
\begin{verbatim}
    double Alpha_s(int Nloop, double scale);
\end{verbatim}
This returns the value of $\alpha_S(\mu)$, evaluated at renormalization scale \texttt{scale}, where the running uses the $\beta$-function evaluated at either one- or two-loop order (\texttt{Nloop}=1 or 2 respectively).
\begin{verbatim}
    double Alpha_em(double scale);
\end{verbatim}
This returns the value of $\alpha_{{\rm EM}}(\mu)$, evaluated at renormalization scale \texttt{scale}. The running of $\alpha_{{\rm EM}}$ with scale is not implemented at present, this method simply returns the value of the coupling at a renormalization scale equal to the $Z$ boson mass, i.e.\ $\alpha_{{\rm EM}}(M_Z)$. However since this coupling is always accessed through this method, the running of the coupling could easily be included if needed.

The following `get' methods may be used to obtain the values of various parameters.
\begin{verbatim}
    double getUpMass();
\end{verbatim}
This returns the up quark mass.
\begin{verbatim}
    double getDownMass();
\end{verbatim}
This returns the down quark mass.
\begin{verbatim}
    double getCharmMass();
\end{verbatim}
This returns the charm quark mass.
\begin{verbatim}
    double getStrangeMass();
\end{verbatim}
This returns the strange quark mass.
\begin{verbatim}
    double getTopMass();
\end{verbatim}
This returns the top quark mass.
\begin{verbatim}
    double getBottomMass();
\end{verbatim}
This returns the bottom quark mass.
\begin{verbatim}
    double getTopWidth();
\end{verbatim}
This returns the top quark width.
\begin{verbatim}
    double getBottomWidth();
\end{verbatim}
This returns the bottom quark width.
\begin{verbatim}
    int    getNoOfLightFlavours();
\end{verbatim}
This returns the number of light quark flavours.
\begin{verbatim}
    double getWmass();
\end{verbatim}
This returns the $W$ boson mass.
\begin{verbatim}
    double getZmass();
\end{verbatim}
This returns the $Z$ boson mass.
\begin{verbatim}
    double getHmass();
\end{verbatim}
This returns the Higgs boson mass.
\begin{verbatim}
    double getWwidth();
\end{verbatim}
This returns the $W$ boson width.
\begin{verbatim}
    double getZwidth();
\end{verbatim}
This returns the $Z$ boson width.
\begin{verbatim}
    double getHwidth();
\end{verbatim}
This returns the Higgs boson width.
\begin{verbatim}
    double getSin2W();
\end{verbatim}
This returns the value of $\sin^2\theta_W$.
\begin{verbatim}
    double getGfermi();
\end{verbatim}
This returns the value of Fermi's constant, $G_F$.
\begin{verbatim}
    double getAlphaMZ();
\end{verbatim}
This returns the value of the electromagnetic coupling at the $Z$ boson mass, $\alpha_{{\rm EM}}(M_Z)$.
\begin{verbatim}
    double getAlphasMZ();
\end{verbatim}
This returns the value of the strong coupling at the $Z$ boson mass, $\alpha_{S}(M_Z)$.
\begin{verbatim}
    int    getNc();
\end{verbatim}
This returns the number of colour states of quarks, $N_c$.
\begin{verbatim}
    double getCF();
\end{verbatim}
This returns the value of $C_F$.
\begin{verbatim}
    double getCA();
\end{verbatim}
This returns the value of $C_A$.
\begin{verbatim}
    double getTR();
\end{verbatim}
This returns the value of $T_R$.

The parameters used by \texttt{TeVJet} can be altered by the user, note however that these parameters cannot be changed during the run, and should be set in the main program before the constructors for any of the \texttt{TeVJet} classes are called. This is because these constructors may need access to the SM parameters. For this reason there is a method
\begin{verbatim}
    void Freeze();
\end{verbatim}
which `freezes' the parameters stored in the \texttt{Tools} class so that subsequent calls to any of the `set' methods will not alter the parameters and will print a warning message. This is called by the \texttt{JetFunction} and \texttt{Process} constructors.

The default value for each of the parameters is listed in Table~\ref{default}. The following `set' methods may be used to set the values of these parameters.

\begin{table}[t]
\begin{center}
\begin{tabular}{|l|l|l|r@{~}l|}\hline
Parameter&Name&Type&Default&\\\hline
$m_u$&\texttt{m\_UpMass}&\texttt{double}&0.0&GeV\\
$m_d$&\texttt{m\_DownMass}&\texttt{double}&0.0&GeV\\
$m_c$&\texttt{m\_CharmMass}&\texttt{double}&1.25&GeV\\
$m_s$&\texttt{m\_StrangeMass}&\texttt{double}&0.0&GeV\\
$m_t$&\texttt{m\_TopMass}&\texttt{double}&174.3&GeV\\
$m_b$&\texttt{m\_BottomMass}&\texttt{double}&4.7&GeV\\
$\Gamma_t$&\texttt{m\_TopWidth}&\texttt{double}&1.508&GeV\\
$\Gamma_b$&\texttt{m\_BottomWidth}&\texttt{double}&0.0&GeV\\
$N_f$&\texttt{m\_NoOfLightFlavours}&\texttt{int}&5&\\
EW scheme&\texttt{m\_EWconvention}&\texttt{int}&3&\\
$\alpha_{{\rm EM}}(M_Z)$&\texttt{m\_alpha}&\texttt{double}&1/128.89&\\
$G_F$&\texttt{m\_Gfermi}&\texttt{double}&$1.16639\times 10^{-5}$&$\mathrm{GeV}^{-2}$\\
$\sin^2\theta_W$&\texttt{m\_sin2thetaW}&\texttt{double}&0.231&\\
$M_W$&\texttt{m\_Wmass}&\texttt{double}&80.41&GeV\\
$M_Z$&\texttt{m\_Zmass}&\texttt{double}&91.188&GeV\\
$M_H$&\texttt{m\_Hmass}&\texttt{double}&120.0&GeV\\
$\Gamma_W$&\texttt{m\_Wwidth}&\texttt{double}&2.124&GeV\\
$\Gamma_Z$&\texttt{m\_Zwidth}&\texttt{double}&2.4952&GeV\\
$m_{\tau}$&\texttt{m\_TauMass}&\texttt{double}&1.777&GeV\\
$\alpha_{S}(M_Z)$&\texttt{m\_alpha\_s\_MZ}&\texttt{double}&0.118&\\
$N_c$&\texttt{m\_Nc}&\texttt{int}&3&\\
$T_R$&\texttt{m\_TR}&\texttt{double}&0.5&\\\hline
\end{tabular}
\caption{Default values for the SM parameters in \texttt{TeVJet}.}
\label{default}
\end{center}
\end{table}
\begin{verbatim}
    void setUpMass(double mass);
\end{verbatim}
Sets the up quark mass to \texttt{mass}.
\begin{verbatim}
    void setDownMass(double mass);
\end{verbatim}
Sets the down quark mass to \texttt{mass}.
\begin{verbatim}
    void setCharmMass(double mass);
\end{verbatim}
Sets the charm quark mass to \texttt{mass}.
\begin{verbatim}
    void setStrangeMass(double mass);
\end{verbatim}
Sets the strange quark mass to \texttt{mass}.
\begin{verbatim}
    void setTopMass(double mass);
\end{verbatim}
Sets the top quark mass to \texttt{mass}.
\begin{verbatim}
    void setBottomMass(double mass);
\end{verbatim}
Sets the bottom quark mass to \texttt{mass}.
\begin{verbatim}
    void setTopWidth(double width);
\end{verbatim}
Sets the top quark width to \texttt{width}.
\begin{verbatim}
    void setBottomWidth(double width);
\end{verbatim}
Sets the bottom quark width to \texttt{width}.
\begin{verbatim}
    void setNoOfLightFlavours(int N);
\end{verbatim}
Sets the number of light quark flavours to \texttt{N}.

The electroweak parameters require a little explanation. There are 6 parameters, which can be chosen to be the SU(2) coupling strength $g$, the weak mixing angle $\sin^2\theta_W$, the electromagnetic coupling constant $\alpha_{{\rm EM}}$ and the masses of the $W$, $Z$ and $H$ bosons. However tree-level gauge invariance imposes two constraints such that there are only 4 free parameters, and as such one should not allow all six parameters to be set independently. \texttt{TeVJet} follows the convention in ALPGEN~\cite{Mangano:2002ea}, which does the following. It treats the Higgs mass as a free parameter, it then takes 3 input values, which can be set to the best fit to experimental data and calculates all other parameters from these using the constraints from tree-level gauge invariance. There is a choice of which 3 parameters are used as the input, which is controlled by an integer variable \texttt{m\_EWconvention} and can be set via the method
\begin{verbatim}
   void setEWconvention(int i);
\end{verbatim}
This is equivalent to the variable \texttt{iewopt} in~\cite{Mangano:2002ea}; the possible values are 0, 1, 2 or 3. Table~\ref{alp} lists the input parameters for the different values of \texttt{m\_EWconvention}. 
\begin{table}[t]
\begin{center}
\begin{tabular}{|c|l|}\hline
\texttt{m\_EWconvention}&Input parameters\\ \hline
0 & $\alpha_{{\rm EM}}(M_Z)$, $G_F$, $\sin^2\theta_W$\\ 
1 & $M_W$, $G_F$, $\sin^2\theta_W$\\ 
2 & $M_Z$, $\alpha_{{\rm EM}}(M_Z)$, $\sin^2\theta_W$\\
3 & $M_Z$, $M_W$, $G_F$\\ \hline
\end{tabular}
\caption{The input parameters that can be set by the user for different \texttt{m\_EWconvention} values. The defaults for these parameters are listed in Table~\ref{default}. Note the default values are only used for the input parameters and the remaining parameters are calculated from these.}
\label{alp}
\end{center}
\end{table}

The Higgs boson mass can be set via the method
\begin{verbatim}
    void setHmass(double mass);
\end{verbatim}
which sets the Higgs boson mass to \texttt{mass}.

The following methods can be used to set the other 3 input parameters.
\begin{verbatim}
    void setWmass(double mass);
\end{verbatim}
Sets the $W$ boson mass to \texttt{mass}.
\begin{verbatim}
    void setZmass(double mass);
\end{verbatim}
Sets the $Z$ boson mass to \texttt{mass}.
\begin{verbatim}
    void setSin2W(double sin2w);
\end{verbatim}
Sets the value of $\sin^2\theta_W$ to \texttt{sin2w}.
\begin{verbatim}
    void setGfermi(double Gfermi);
\end{verbatim}
Sets the value of Fermi's constant to \texttt{Gfermi}.
\begin{verbatim}
    void setAlphaMZ(double alpha);
\end{verbatim}
Sets the value of the electromagnetic coupling at the $Z$ boson mass, $\alpha_{{\rm EM}}(M_Z)$.

Once one of the three input values for the current \texttt{m\_EWconvention} value is set, the other parameters are automatically recalculated. If one attempts to set one of the parameters that is not an input parameter for the current \texttt{m\_EWconvention} value, \texttt{TeVJet} will print a warning statement, and that parameter will {\bf not} be changed.
\begin{verbatim}
    void setWwidth(double width);
\end{verbatim}
Sets the $W$ boson width to \texttt{width}.
\begin{verbatim}
    void setZwidth(double width);
\end{verbatim}
Sets the $Z$ boson width to \texttt{width}.
\begin{verbatim}
    void setHwidth(double width);
\end{verbatim}
Sets the Higgs boson width to \texttt{width}.
\begin{verbatim}
    void setAlphasMZ(double alphas);
\end{verbatim}
Sets the value of the strong coupling at the $Z$ boson mass, $\alpha_{S}(M_Z)$.
\begin{verbatim}
    void setNc(int N);
\end{verbatim}
Sets the number of colour states of a quark to \texttt{N}. If this is set the values returned by the methods \texttt{getCF()} and \texttt{getCA()} are altered accordingly. Also the QCD $\beta$-function is recalculated and the running of $\alpha_S$ will change.
\begin{verbatim}
    void setTR(double TR);
\end{verbatim}
Sets the value of $T_R$ to \texttt{TR}.

\subsection{The PhaseSpace class}
\label{phase}
There are currently three phase space classes that inherit from the \texttt{PhaseSpace} class:\\ \texttt{TwoBodyPhaseSpace}, \texttt{ThreeBodyPhaseSpace} and \texttt{FourBodyPhaseSpace}. 

An element of massless two-body phase space can be written as
\begin{equation}
d\phi^{(2)}(p_1,p_2;Q)=\frac{1}{8\pi}d\cos\theta^*d\phi,
\end{equation}
where $\theta^*$ and $\phi$ are the polar and azimuthal angles of one of the particles in the centre of mass frame. By default, points in two-body phase space are chosen flat in $\cos\theta^*$  and $\phi$. For processes with a hadron in the initial state the partonic cross section must be folded with the parton distribution functions, so we need to calculate quantities like
\begin{eqnarray}
I&=&\int_0^1d\eta\,f(\eta)\int d\phi^{(2)}|\mathcal{M}|^2\nonumber\\
&=&\frac{1}{8\pi}\int d\eta\,d\cos\theta^*d\phi\,f(\eta)|\mathcal{M}|^2.
\end{eqnarray}
In general it is very inefficient to pick flat in $\eta$.  Instead, we can use the mapping
\begin{equation}
d\eta\,d\cos\theta^*=\frac{4\eta e^{-y}}{\sqrt{s}}\,dp_t\,dy,
\end{equation}
where $p_t$ and $y$ are the transverse momentum and rapidity of one of the particles in the centre of mass frame of the beams. It is possible to pick flat in $y$ and as $p_t^n$ via the method \texttt{SetMode}.

The three-body phase space generator uses a two-body configuration and the phase space factorization of Eq.~(\ref{phase-fact}) to produce one additional particle. In 4 dimensions the dipole phase space factorization is
\begin{eqnarray}
d\phi^{(3)}(p_i,p_j,p_k;Q)&=&\frac{\widetilde{p}_{ij}.\widetilde{p}_k}{16\pi^3}(1-y_{ij,k})\,\Theta(\widetilde{z}_i(1-\widetilde{z}_i))\,\Theta(y_{ij,k}(1-y_{ij,k}))\,d\widetilde{z}_i\,dy_{ij,k}\,d\phi\nonumber\\
&&\times d\phi^{(2)}(\widetilde{p}_{ij},\widetilde{p}_k;Q).
\end{eqnarray}
Since this is an exact factorization, in order to integrate a well-behaved function of phase space it would be sufficient to follow the following algorithm to produce three-particle final states:
\begin{enumerate}
\item Pick a two-body configuration ($\widetilde{p}_{12},\widetilde{p}_3$) with weight $w_2$
\item Pick the dipole variables $\widetilde{z}_i,$ $y_{ij,k}$ and $\phi$ in the appropriate ranges
\item Use these and the two-body configuration to reconstruct the three-body configuration ($p_1,p_2,p_3$) where $\widetilde{p}_{12}$ is the emitter and $\widetilde{p}_3$ the spectator. This should be given weight
\begin{equation}
w_3=w_2\times\frac{\widetilde{p}_{12}.\widetilde{p}_3}{16\pi^3}(1-y_{12,3}).
\end{equation}
\end{enumerate}
However the integration of the real matrix element minus the subtraction terms over phase space typically contains square root singularities in the dipole variables, i.e.
\begin{eqnarray}
\left(|\mathcal{M}_{m+1}|^2F_J^{(m+1)}(p)-\sum_{{\rm dipoles}}(\mathcal{D}.F_J^{(m)})(p)\right)\longrightarrow \frac{1}{\sqrt{y_{ij,k}}},\qquad y_{ij,k}\rightarrow 0
\end{eqnarray}
and
\begin{eqnarray}
\left(|\mathcal{M}_{m+1}|^2F_J^{(m+1)}(p)-\sum_{{\rm dipoles}}(\mathcal{D}.F_J^{(m)})(p)\right)\longrightarrow 
\left\{\begin{array}{cc}
\frac{1}{\sqrt{\widetilde{z}_i}} & \widetilde{z}_i\rightarrow 0,\\
\frac{1}{\sqrt{1-\widetilde{z}_i}} & \widetilde{z}_i\rightarrow 1.
\end{array}\right.
\end{eqnarray}
Therefore in order for the Monte Carlo integration to converge one must introduce a mapping to remove these singularities. In order to do this we introduce a Jacobian factor of the form:
\begin{eqnarray}
\mathcal{J}=\frac{8\sqrt{y_{ij,k}}\sqrt{\widetilde{z}_i(1-\widetilde{z}_i)}}{\sqrt{\widetilde{z}_i}+\sqrt{1-\widetilde{z}_i}},
\label{jac}
\end{eqnarray}
i.e.\ the dipole variables are chosen such that the weight should be multiplied by this factor, which tends to zero as any of the limits $y_{ij,k}\rightarrow 0$ or $\widetilde{z}_i\rightarrow 0,1$ are approached. Since the integrand contains square root singularities in these limits for {\bf any} choice of $i,j$ and $k$ \texttt{TeVJet} uses a multichannel phase space generation so that instead of fixing the choice of emitter and spectator all possible choices are used. Each channel is chosen with equal probability, and the total Jacobian factor $\mathcal{J}$ is given by
\begin{equation}
\frac{1}{\mathcal{J}}=\sum_i\frac{1}{\mathcal{J}_i},
\end{equation}
where $\mathcal{J}_i$ is the Jacobian factor for the $i$th channel, given by Eq.~(\ref{jac}).

Four-body configurations are generated from three-body configurations in a similar way.

\subsection*{PhaseSpace methods}
The constructor for the \texttt{PhaseSpace} classes takes a \texttt{std::vector} with the masses of the final state particles. 
\begin{verbatim}
    PhaseSpace(std::vector<double> massvec);
\end{verbatim}
At present only massless phase space generation is implemented, so \texttt{massvec} must be filled with zeros, but we anticipate the inclusion of masses soon. The way that two-body phase space is picked can be altered via the method:
\begin{verbatim}
    void SetMode(std::string mode);
\end{verbatim}
Possible values of \texttt{mode} are \texttt{lepton-lepton} where two-body phase space is picked flat in $\theta^*$ and $\phi$; \texttt{DIS} where two-body phase space is picked flat in the rapidity $y$ and as some power  of the $p_t$ ($p_t^{{\rm ptpow}}$) of one of the outgoing particles in the centre of mass frame of the beams; or \texttt{hadron-hadron} where two-body phase space is picked flat in the rapidities of the two outgoing particles, $y_{1,2}$, and the transverse momentum of the particles to some power, $p_t^{{\rm ptpow}}$, in the centre of mass frame of the beams.

When the mode has been set to \texttt{DIS} the user may choose the minimum $p_t$ and the power ptpow via the methods
\begin{verbatim}
    void SetPtpow(double ptpow);
    void SetPtmin(double ptmin);
\end{verbatim}

\subsection{The Histogram class}
\label{histo1}
The \texttt{Histogram} class can be used to store the distributions calculated by the \texttt{JetFunction} class. Two versions of the \texttt{Histogram} class source code are available: \texttt{HistogramSimple.cc} and \texttt{HistogramRoot.cc}. The `simple' version is a very basic histogram package and is used by default, the other version is a wrapper for the Root~\cite{Brun:1997pa} class \texttt{TH1F}. In order to use the Root version one must remove \texttt{HistogramSimple.cc} from the source directory and replace it by the Root version, which is in the directory \texttt{Histcode/}; uncomment the relevant lines in the header file (\texttt{Histogram.h}); and uncomment the lines in the Makefile that link to the Root libraries. If the user wishes to use another histogram package they should write the necessary source code for the methods described below.
\subsection*{Histogram methods}
The constructor takes the name and title of the histogram, the number of bins, and the minimum and maximum values of the quantity:
\begin{verbatim}
    Histogram(std::string name, std::string title,
              int Nbins, double min, double max);
\end{verbatim}
The following methods are available to fill quantity $x$ either unweighted (i.e.\ with weight 1) or with weight \texttt{weight}:
\begin{verbatim}
    void   Fill(double x);
    void   Fill(double x, double weight);
\end{verbatim}
If the following method is called, the \texttt{Histogram} will store the error on each bin by adding the weights in quadrature:
\begin{verbatim}
    void   Sumw2();
\end{verbatim}
The following methods return the width, error, centre or content of bin number \texttt{bin} (Note: the bins are labeled from 1 to \texttt{Nbins}):
\begin{verbatim}
    double GetBinWidth(int bin);
    double GetBinError(int bin);
    double GetBinCenter(int bin);
    double GetBinContent(int bin);
\end{verbatim}
The following method returns the number of bins:
\begin{verbatim}
    int    GetNbinsX();
\end{verbatim}
The following methods return the minimum or maximum values of the observable:
\begin{verbatim}
    double GetMinimum();
    double GetMaximum();
\end{verbatim}
The following method returns the title of the histogram:
\begin{verbatim}
    std::string GetTitle();
\end{verbatim}

\subsection{The HistoFile class}
\label{histo2}
The \texttt{HistoFile} class is the file that the \texttt{Histogram}s stored in the vector \texttt{m\_histovec} of the \texttt{JetFunction} class are written to at the end of the run. Two versions of the HistoFile class source code are available: \texttt{HistoFileSimple.cc} and \texttt{HistoFileRoot.cc}. The `simple' version is a text file that contains names of each histogram and the values and errors of each bin and is used by default. The other version is a wrapper for the Root~\cite{Brun:1997pa} class \texttt{TFile}. In order to use the Root version one must remove \texttt{HistoFileSimple.cc} from the source directory and replace it by the Root version, which is in the directory \texttt{Histcode/}; uncomment the relevant lines in the header file (\texttt{HistoFile.h}); and uncomment the lines in the Makefile that link to the Root libraries. If the user wishes to use another file format they should write the necessary source code for the methods described below.

\subsection*{HistoFile methods}
The constructor takes a \texttt{std::string}, which is the name of the output file. This will be appended with .simple or .root depending on which version of \texttt{HistoFile.cc} is used:
\begin{verbatim}
    HistoFile(std::string fname);
\end{verbatim}
The following method writes the \texttt{Histogram}s in \texttt{histos} to the output file:
\begin{verbatim}
    void Write(std::vector<Histogram*> histos);
\end{verbatim}
The following method closes the output file:
\begin{verbatim}
    void Close();
\end{verbatim}

\subsection{The MatrixElementSquared class}
\label{mat}
The \texttt{MatrixElementSquared} class is an abstract base class. The mechanism for providing code to calculate the square of the matrix element for a sub-process that contributes to a process is to make a class that inherits from the \texttt{MatrixElementSquared} class. Pointers to instances of this then get `pushed back' into one of the vector members of the \texttt{Process} in its \texttt{Initialize} method. When writing a class that inherits from the \texttt{MatrixElementSquared} class, the user must initialize the following members:
\begin{verbatim}
    int m_NInitColour;
\end{verbatim}
This is the number of initial state colour states, e.g.\ for the process $qg\rightarrow qg$ this would be $3\times 8=24$.
\begin{verbatim}
    int m_BoseSymFactor;
\end{verbatim}
This is the Bose symmetry factor that one must divide by when there are identical particles in the final state.

In addition to these, the following members must be initialized:
\begin{verbatim}
    std::vector<int> m_InitialID;
    std::vector<int> m_FinalID;
    std::vector<int> m_IDlist;
\end{verbatim}
These are initialized in the methods \texttt{SetParticleIDs} and \texttt{FillIDlist} (see below).
\subsubsection*{MatrixElementSquared methods}
When writing a class that inherits from the \texttt{MatrixElementSquared} class, the following virtual methods must be written.
\begin{verbatim}
    virtual void SetParticleIDs()=0;
\end{verbatim}
In this method the user must push back the PDG ID codes of the two initial state particles into the member object \texttt{m\_InitialID} and the ID codes of all final state particles into the member object \texttt{m\_FinalID}. Finally the user should call the method \texttt{FillIDlist}. Note: this method {\bf must} be called by the constructor of this class.
\begin{verbatim}
    virtual double GetMSquared(
                    HepLorentzVector part1, 
                    HepLorentzVector part2, 
                    std::vector<HepLorentzVector> finalstate, 
                    double RenScale)=0;
\end{verbatim}
This is a method that should return the square of the matrix element given the initial and final state momenta, and the renormalization scale. The momenta will be passed in the order set by the user in the method \texttt{SetParticleIDs}.
\subsection{The Amplitude class}
\label{amp}
The \texttt{Amplitude} class is an abstract base class, which has a number of members, which must be initialized and virtual methods, which must be overwritten.

It has been shown~\cite{Maltoni:2002mq} that QCD amplitudes can be decomposed (the colour-flow decomposition) in terms of different colour structures as
\begin{equation}
\mathcal{M}=\sum_{a}c_a\mathcal{A}_a,
\label{decom}
\end{equation}
where the factor associated with each colour structure, $\mathcal{A}_a$, is called a partial amplitude. Thus the square of the matrix element for a scattering process can be written as
\begin{eqnarray}
|\mathcal{M}|^2&=&\sum_{a,b}c_a\mathcal{A}_ac_b^*\mathcal{A}_b^*\nonumber\\
&=&\sum_{a,b}C_{ab}Q_{ab},
\label{colour-flow}
\end{eqnarray} 
where the elements of the colour flow matrix are given by $C_{ab}=c_ac_b^*$ and the `quark matrix' is given by $Q_{ab}=\mathcal{A}_a\mathcal{A}_b^*$.
Many of the terms needed in the dipole subtraction method are proportional to the square of the colour correlated Born level matrix element given by a similar expression, but with different colour matrices:
\begin{eqnarray}
|\mathcal{M}^{ij}|^2=\sum_{a,b}\widetilde{C}_{ab}^{ij}Q_{ab}.
\end{eqnarray} 
Thus the \texttt{Calculation} class needs access to both
\begin{itemize}
\item the `quark matrix', $Q_{ab}$, and
\item the set of colour matrices, $\{\widetilde{C}_{ab}^{ij}\}$.
\end{itemize}
In fact since we are usually interested in summing/averaging over final-/initial-state spins, the \texttt{Calculation} class requires the object
\begin{eqnarray}
\overline{Q}_{ab}=\sum_{\{s\}}Q_{ab},
\end{eqnarray} 
where the sum is over the set of helicity configurations of the external particles. In addition to this, in the case of a gluon emitter, the subtraction terms have spin correlations. For scattering processes with an external gluon, each partial amplitude can be written as:
\begin{equation}
\mathcal{A}_a=\epsilon^*_{\mu}\mathcal{A}_a^{\mu},
\end{equation}
where $\epsilon^*_{\mu}$ is the polarization vector of the external gluon. The spin averaged quark matrix is then given by:
\begin{eqnarray}
\overline{Q}_{ab}&=&\sum_{\{s\}}\mathcal{A}_a\mathcal{A}_b^*\nonumber\\
&=&\sum_{\{s\}}\epsilon^{s_g *}_{\mu}\mathcal{A}_a^{\mu}\epsilon^{s_g}_{\nu}\mathcal{A}_b^{\nu *}\nonumber\\
&=&\sum_{\{s^{\prime}\}}\sum_{s_g=1,2}\epsilon^{s_g *}_{\mu}\epsilon^{s_g}_{\nu}\mathcal{A}_a^{\mu}\mathcal{A}_b^{\nu *},
\end{eqnarray}
where in the last line the first sum is over the set of helicity configurations of all external particles except the gluon. To obtain the spin averaged quark matrix one would do these sums, the second of which can be done by the replacement
\begin{equation}
\sum_{s_g=1,2}\epsilon^{s_g *}_{\mu}\epsilon^{s_g}_{\nu}\rightarrow -g_{\mu\nu}.
\end{equation}
In the case of a gluon emitter, the subtraction term  is found by the replacement
\begin{equation}
\sum_{s_g=1,2}\epsilon^{s_g *}_{\mu}\epsilon^{s_g}_{\nu}\rightarrow V_{\mu\nu}.
\end{equation}
Thus the \texttt{Calculation} class requires access to:
\begin{itemize}
\item Each partial amplitude projected onto the helicity space of each external gluon, $\mathcal{A}_a^{\mu}$.
\end{itemize}

When writing a class that inherits from the \texttt{amplitude} class, the user must initialize the following members:
\begin{verbatim}
    std::vector<int> m_InitialID;
    std::vector<int> m_FinalID;
    std::vector<int> m_IDlist;
\end{verbatim}
These are initialized in the methods \texttt{SetParticleIDs} and \texttt{FillIDlist} (see below).
\begin{verbatim}
    unsigned int m_Nhel;
\end{verbatim}
When calculating the subtraction term with a gluon emitter, the \texttt{Calculation} class explicitly performs the sum/average over final/initial state spins. To calculate the subtraction term with a particular gluon emitter, it needs to know the total number of helicity configurations of the other external particles, this is \texttt{m\_Nhel}.
\begin{verbatim}
    std::vector<std::vector<int> > m_HelicityVec;
\end{verbatim}
It also needs to know each of these helicity configurations. This object stores all of these.  
\begin{verbatim}
    unsigned int m_NinStates;
\end{verbatim}
This is the total number of initial state colour and spin states for the sub-process. For example for the process $qg\rightarrow qg$ there are $2\times 2=4$ spin states and $3\times 8=24$ colour states, so \texttt{m\_NinStates} should be set to $4\times 24=96$. 
\begin{verbatim}
    unsigned int m_Ncolour;
\end{verbatim}
This is the number of terms in the decomposition of Eq.~(\ref{decom}), i.e.\ the number of different colour structures.
\begin{verbatim}
    HepMatrix    m_matrix;
\end{verbatim}
This is the `colour flow matrix' of Eq.~(\ref{colour-flow}).

\subsubsection*{Amplitude methods}
When writing a class that inherits from the \texttt{Amplitude} class, the following virtual methods must be over-written. 

The constructor of the class that inherits from the \texttt{Amplitude} class {\bf must} call the first three methods: \texttt{SetParticleIDs}, \texttt{SetHelicityConfigs} and \texttt{InitColourMatrices}.
\begin{verbatim}
    virtual void SetParticleIDs();
\end{verbatim}
In this method the user must push back the PDG ID codes of the two initial state particles into the member object \texttt{m\_InitialID} and the ID codes of all final state particles into the member object \texttt{m\_FinalID}. Finally the user should call the method \texttt{FillIDlist}.
\begin{verbatim}
    virtual void SetHelicityConfigs();
\end{verbatim}
When calculating the subtraction term with a gluon emitter, the \texttt{Calculation} class explicitly performs the sum/average over final/initial state spins. To calculate the subtraction term with a particular gluon emitter, it needs access to each partial amplitude projected onto the helicity space of that gluon for {\bf each} helicity configuration of the other external particles. The two helicity states of quarks and gluons are both stored as $\pm 1$. In this method, the user should `push back' into the vector \texttt{m\_HelicityVec} a set of vectors of integers (either $\pm 1$), which represent the helicity configurations of the other external particles. 
\begin{verbatim}
    virtual void InitColourMatrices();
\end{verbatim}
Here the user should initialize the elements of the colour flow matrix, \texttt{m\_matrix}. The user should also initialize the set of colour flow matrixes, $\{\widetilde{C}_{ab}^{ij}\}$, which are needed for the subtraction terms.
\begin{verbatim}
    virtual HepMatrix GetAmplitudeMatrixQuark(
                          std::vector<HepLorentzVector> momenta, 
                          double RenScale)=0;
\end{verbatim}
This returns the `quark matrix', $Q_{ab}$, of Eq.~(\ref{colour-flow}).
\begin{verbatim}
    virtual complex<HepLorentzVector> GetAmplitudeGluon(
                          std::vector<HepLorentzVector> momenta, 
                          double RenScale, 
                          int ijtilda, int Ncol, int ihel)=0;
\end{verbatim}
This should return the \texttt{Ncol}-th partial amplitude projected onto the helicity space of the gluon, which is the \texttt{ijtida}-th element of \texttt{m\_IDlist}. The other external helicities are given by the \texttt{ihel}-th element of \texttt{m\_HelicityVec}.
\begin{verbatim}
    virtual HepMatrix GetColourMatrix(int ijtilda, int ktilda)=0;
\end{verbatim}
This returns the colour flow matrix $\widetilde{C}^{\widetilde{ij} \widetilde{k}}$.

\subsection{The LoopIntegral class}
\label{loop}
The \texttt{LoopIntegral} class is an abstract base class. The mechanism for providing code to calculate the interference between the Born level and 1-loop amplitudes for a sub-process that contributes to a process is to make a class that inherits from the \texttt{LoopIntegral} class. Pointers to instances of all such classes then get `pushed back' into the vector in the \texttt{Initialize} method of the \texttt{Process} class. 

\subsubsection*{LoopIntegral methods}
The \texttt{LoopIntegral} class has one virtual method, which must be over-written:
\begin{verbatim}
    virtual double GetLoop(HepLorentzVector part1, 
                           HepLorentzVector part2, 
                           std::vector<HepLorentzVector> finalstate, 
                           double RenScale)=0;
\end{verbatim}
The user should provide code that returns the finite part of the renormalized one-loop matrix element. This should be calculated in the $\overline{\mathrm{MS}}$ scheme. What we mean by `the finite part' of the renormalized one-loop matrix element is explained in section~\ref{end}. Using dimensional regularization the one-loop matrix element in the $\overline{\mathrm{MS}}$ can be written in the form
\begin{equation}
|\mathcal{M}|^2_{1-loop}=\frac{\alpha_S}{2\pi}\frac{1}{\Gamma(1-\epsilon)}\left(\frac{4\pi\mu^2}{Q^2}\right)^{\epsilon}\left\{\frac{A}{\epsilon^2}+\frac{B}{\epsilon}+C+{\cal O}(\epsilon)\right\},
\end{equation}
where $\mu$ is the renormalization scale, and $Q$ is some scale relevant to the process. The user should provide code that returns 
\begin{equation}
\frac{\alpha_S}{2\pi}C.
\end{equation}
The scale $Q$ {\bf must} be the scale that is returned by the \texttt{getRefScale} method of the \texttt{Process} class (see section~\ref{proc}).
\section{Using \texttt{TeVJet}}
\label{use}
\subsection{Installing \texttt{TeVJet}}
The \texttt{TeVJet} source code is available from:
\begin{quote}\texttt{%
http://www.hep.man.ac.uk/u/chris/tevjet 
}\end{quote}
To unpack this type:
\begin{quote}\texttt{%
tar -zxvf TeVJet-1.0.0.tar.gz
}\end{quote}
This will create a directory \texttt{TeVJet-1.0.0}. \texttt{TeVJet} must be linked at compilation time to HELAS~\cite{Murayama:1992gi}, CLHEP~\cite{Lonnblad:1994kt} (version 1.8 only at present) and LHAPDF~\cite{lhapdf}. The source code for the HELAS library is distributed with \texttt{TeVJet}, and can be compiled by typing (in the directory \texttt{TeVJet-1.0.0})
\begin{quote}\texttt{%
make libdhelas3
}\end{quote}
CLHEP (version 1.8) and LHAPDF must both be installed\footnote{They can be obtained from the web addresses given in Refs.~\cite{Lonnblad:1994kt,lhapdf} together with installation instructions.} and the user must edit \texttt{TeVJet}'s Makefile to provide the path to the libraries. Once this is done type:
\begin{quote}\texttt{%
make
}\end{quote}
This will build the \texttt{TeVJet} library and some example programs. For processes with initial state hadrons \texttt{TeVJet} needs access to the LHAPDF grid files type:
\begin{quote}\texttt{%
make PDFsets
}\end{quote}
This will create a symbolic link to the grid files.

One of the example programs that will have been built is called \texttt{EEto3jet}. We show here its main program and the output that it produces.
\subsection{Example main program}
\label{main.cc}
\begin{verbatim}
//standard lib headers
#include <iostream>
#include <cmath>

//my lib headers that are always needed
#include "Tools.h"
#include "Calculation.h"
#include "JetFunction.h"
#include "Process.h"

//process specific user headers
#include "ProcEEto3jet.h"
#include "UserEEto3jet.h"

int main(int argc, char** argv) {

  long           seed = 12345;
  std::string filename="EEto3jet_12345";

  //set the SM parameters
  TeVJet::Tools::getInstance()->setEWconvention(3);
  TeVJet::Tools::getInstance()->setWmass(80.41);
  TeVJet::Tools::getInstance()->setZmass(91.188);
  TeVJet::Tools::getInstance()->setGfermi(1.16639*pow(10,-5));
  TeVJet::Tools::getInstance()->setAlphasMZ(0.118);

  //set some parameters
  int       Nborn = 1000;//no of 'events'
  int       Nvirt = 1000;
  int       Nreal = 1000;
 
  double CMenergy = TeVJet::Tools::getInstance()->getZmass();//GeV
  bool       useZ = false;
  
  double Ebeam1   = CMenergy/2;
  double Ebeam2   = CMenergy/2;

  double fren = 1.0;

  TeVJet::Process     * proc;
  TeVJet::JetFunction * jet;

  //process specific stuff
  proc = new TeVJet::ProcEEto3jet(CMenergy,useZ);
  jet  = new TeVJet::UserEEto3jet(filename,CMenergy,useZ);
  //process specific ends

  TeVJet::Calculation calc(proc,jet);
  //user can set stuff here
  calc.setNborn(Nborn);
  calc.setNreal(Nreal);
  calc.setNvirt(Nvirt);
  calc.setSeed(seed);
  calc.setEbeam1(Ebeam1);
  calc.setEbeam2(Ebeam2);

  calc.setRenScaleFactor(fren);

  //initialize stuff
  calc.Initialize();

  //calculate born level
  calc.BornLoop();

  //if required calculate NLO contribution
  if(calc.m_proc->DoNLO()){
    calc.RealLoop();
    calc.VirtLoop();
    if(calc.m_proc->HadronBeams()){
      calc.ConvLoop();
    }
  }

  delete jet;
  jet=0;
  delete proc;
  proc=0;

  return 0;
}
\end{verbatim}
\subsection{Example output}
\begin{verbatim}
         
         ----------------------------------------------
         TeVJet Version 1.0.0 - Written by C.Tevlin
         A general framework for calculating jet cross
         sections in NLO QCD. This is an implementaion 
         of the Dipole Subtraction Method of S.Catani & 
         M.H.Seymour.
         (Last Updated 1/2/2008)
         ----------------------------------------------
         
Info: TeVJet::Tools::PrintSMParameters()
      Table of SM parameters used for this run:

---------------------------------------
  UpMass               = 0
  DownMass             = 0
  CharmMass            = 1.25
  StrangeMass          = 0
  TopMass              = 174.3
  BottomMass           = 4.7
  TopWidth             = 1.508
  BottomWidth          = 0
  No of light flavours = 5
  EW scheme            = 3
  alpha_ew(MZ)         = 0.00755099
  Fermi's constant     = 1.16639e-05
  sin^2(thetaW)        = 0.222421
  Wmass                = 80.41
  Zmass                = 91.188
  Hmass                = 120
  Wwidth               = 2.124
  Zwidth               = 2.4952
  TauMass              = 1.777
  alpha_s(MZ)          = 0.118
  Nc                   = 3
  TR                   = 0.5
---------------------------------------

Info: TeVJet::Calculation::Initialize()
      Additional parameters for this run:

---------------------------------------
  Ebeam1               = 45.594
  Ebeam2               = 45.594
  Seed                 = 12345
  Renorm Scheme        = MSbar
  Renorm Scale Factor  = 1
---------------------------------------

Info: Calculating the Born level ...
i = 0
i = 1000
i = 2000
i = 3000
i = 4000
i = 5000
i = 6000
i = 7000
i = 8000
i = 9000

Info: Calculating the Real Diagrams and Subtraction Terms ...
i = 0
i = 1000
i = 2000
i = 3000
i = 4000
i = 5000
i = 6000
i = 7000
i = 8000
i = 9000

Info: Calculating the Virtual + Insertion Operator ...
i = 0
i = 1000
i = 2000
i = 3000
i = 4000
i = 5000
i = 6000
i = 7000
i = 8000
i = 9000

--------------------------------------------------------------
  Average value of (1-T) @ NLO = 0.0558852 +- 0.00224712

    order alpha_s contribution = 0.338854 +- 0.00565262
  order alpha_s^2 contribution = 1.14195 +- 0.154111
--------------------------------------------------------------

\end{verbatim}
By default the only quantity calculated is the average value of 1--thrust, which can be calculated as a perturbative series in $\alpha_s$:
\begin{equation}
\langle1-T\rangle=a_0\alpha_{s}+a_1\alpha_s^2+\dots
\end{equation}
This program prints the full NLO QCD prediction and the two coefficients $a_0$ and $a_1$. We note that this example program is a fairly low statistics run intended for validation purposes, and we have also calculated these quantities with much higher statistics, finding agreement with previous calculations (see for example \cite{Kunszt:1989km}).

The program also produces an output file \texttt{EEto3jet\_12345.simple}, which contains the histograms. This is a text file that is too long to show here, but we note that our version may be found in the directory \texttt{ExampleOutput/}.
\section{Results}
\label{result}
\subsection[$e^+e^-\rightarrow 3$ jets]{\boldmath$e^+e^-\rightarrow 3$ jets}
In this section we present some results for three-jet event shapes at LEP. We have used the one-loop renormalized matrix element from~\cite{Giele:1991vf}. The renormalization scale is set to the centre of mass energy, which is chosen to be equal to the $Z$ boson mass (which is set to 91.188~GeV). We present results for the thrust distribution and the $C$-parameter~\cite{Ellis:1980wv}.

\begin{figure}
\includegraphics[width=0.5\textwidth]{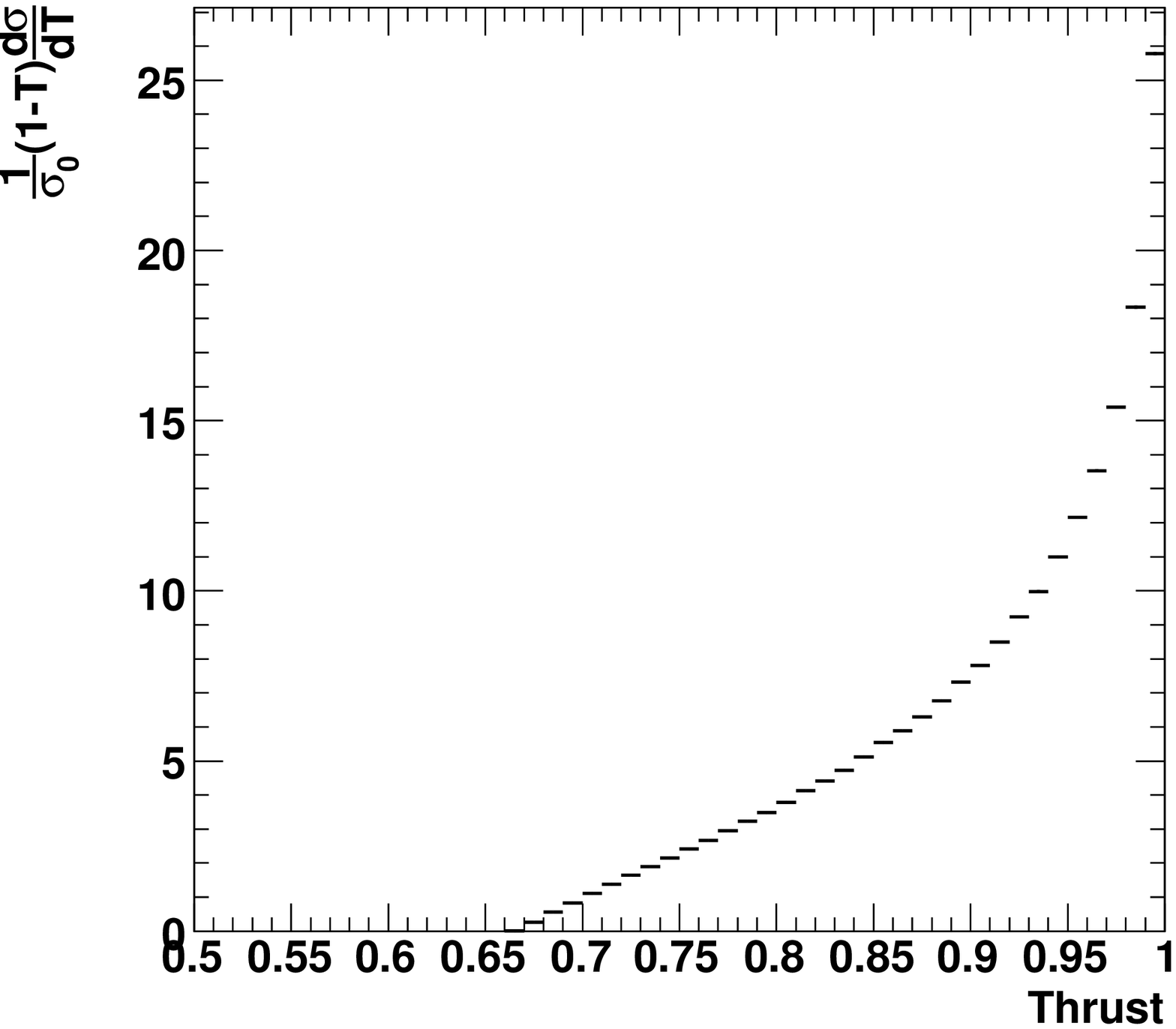}
\includegraphics[width=0.5\textwidth]{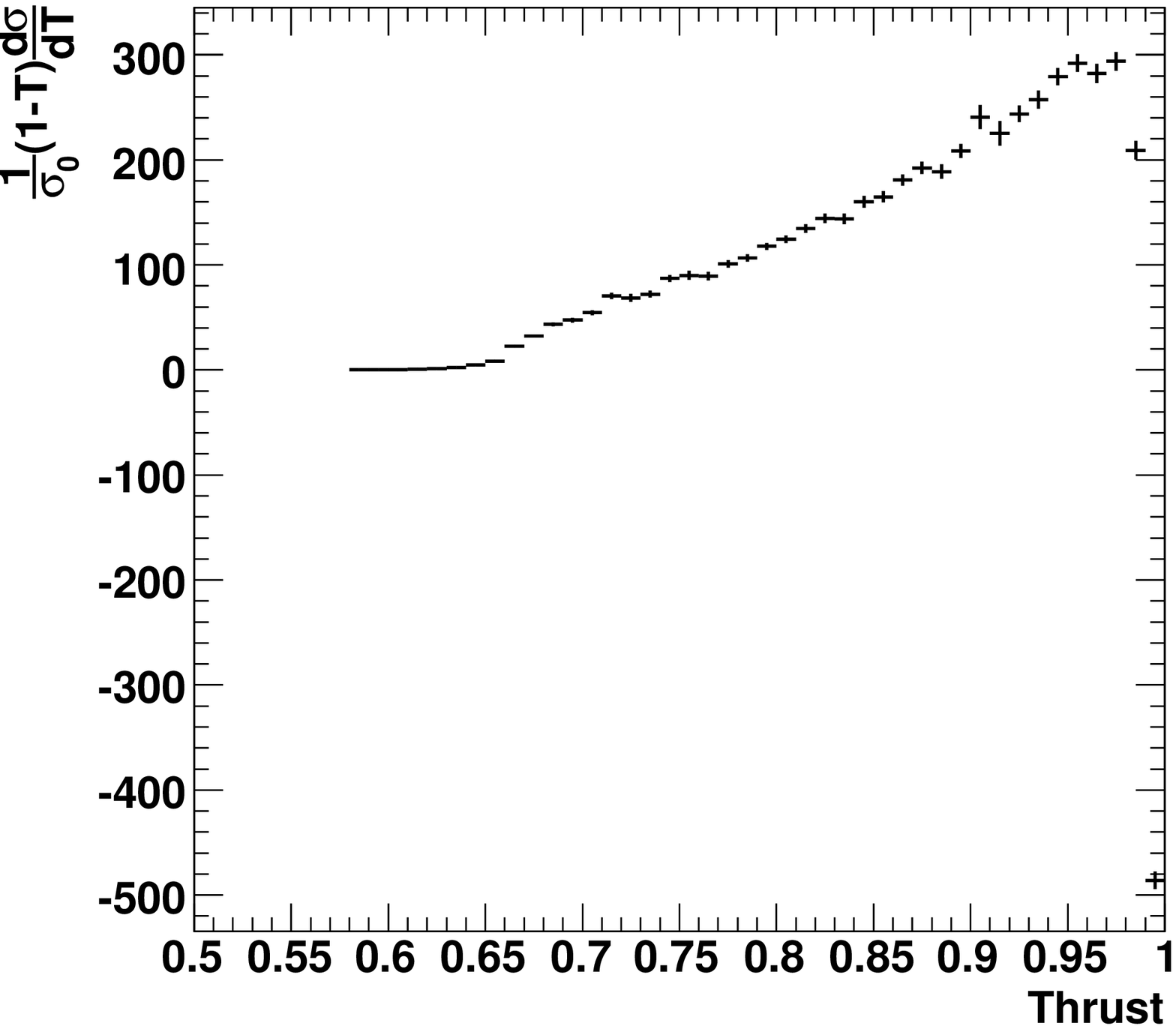}
\caption{The distributions for the thrust coefficient functions (a) $A_t$ and (b) $B_t$.}
\label{t1}
\end{figure}

\begin{figure}
\begin{center}
\includegraphics[width=0.5\textwidth]{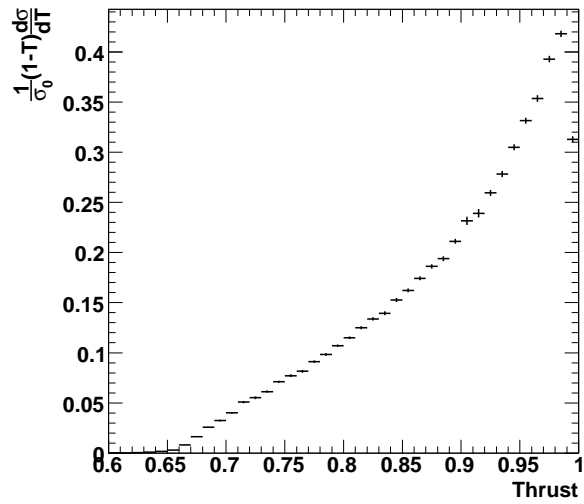}
\caption{The full next to leading order QCD prediction for the thrust distribution.}
\label{t2}
\end{center}
\end{figure}

\begin{figure}
\includegraphics[width=0.5\textwidth]{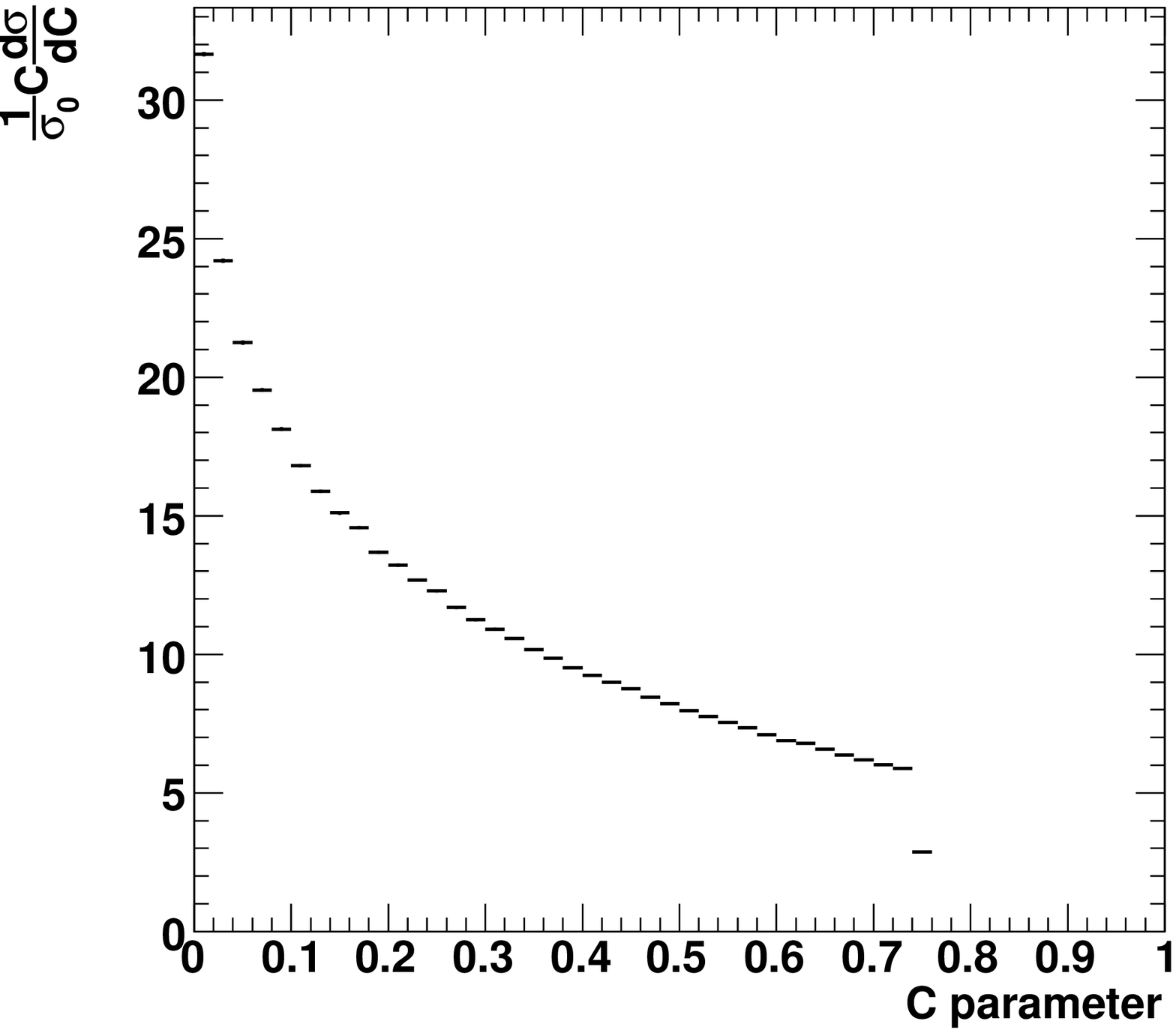}
\includegraphics[width=0.5\textwidth]{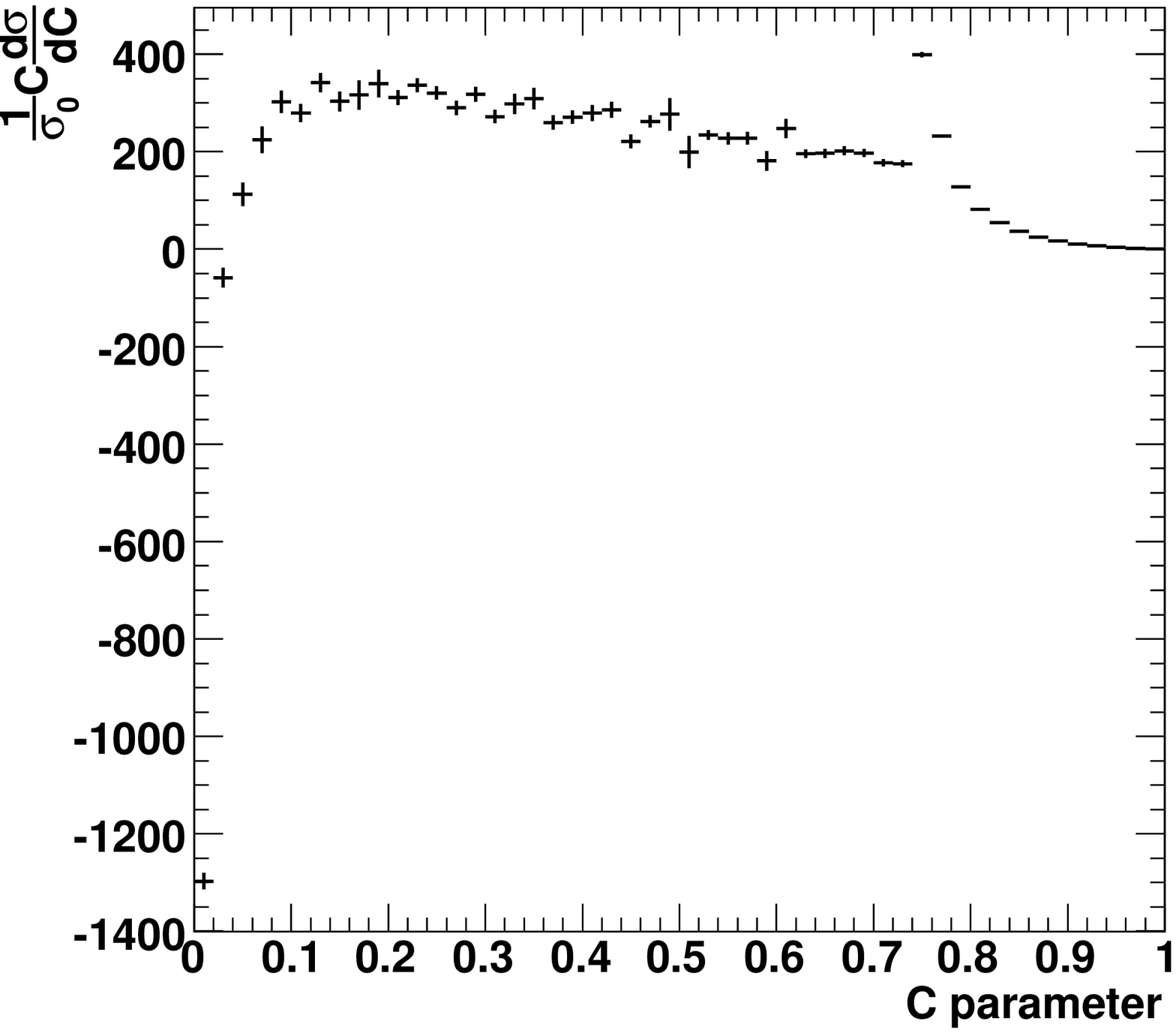}
\caption{The distributions for the thrust coefficient functions (a) $A_C$ and (b) $B_C$.}
\label{c1}
\end{figure}

\begin{figure}
\begin{center}
\includegraphics[width=0.5\textwidth]{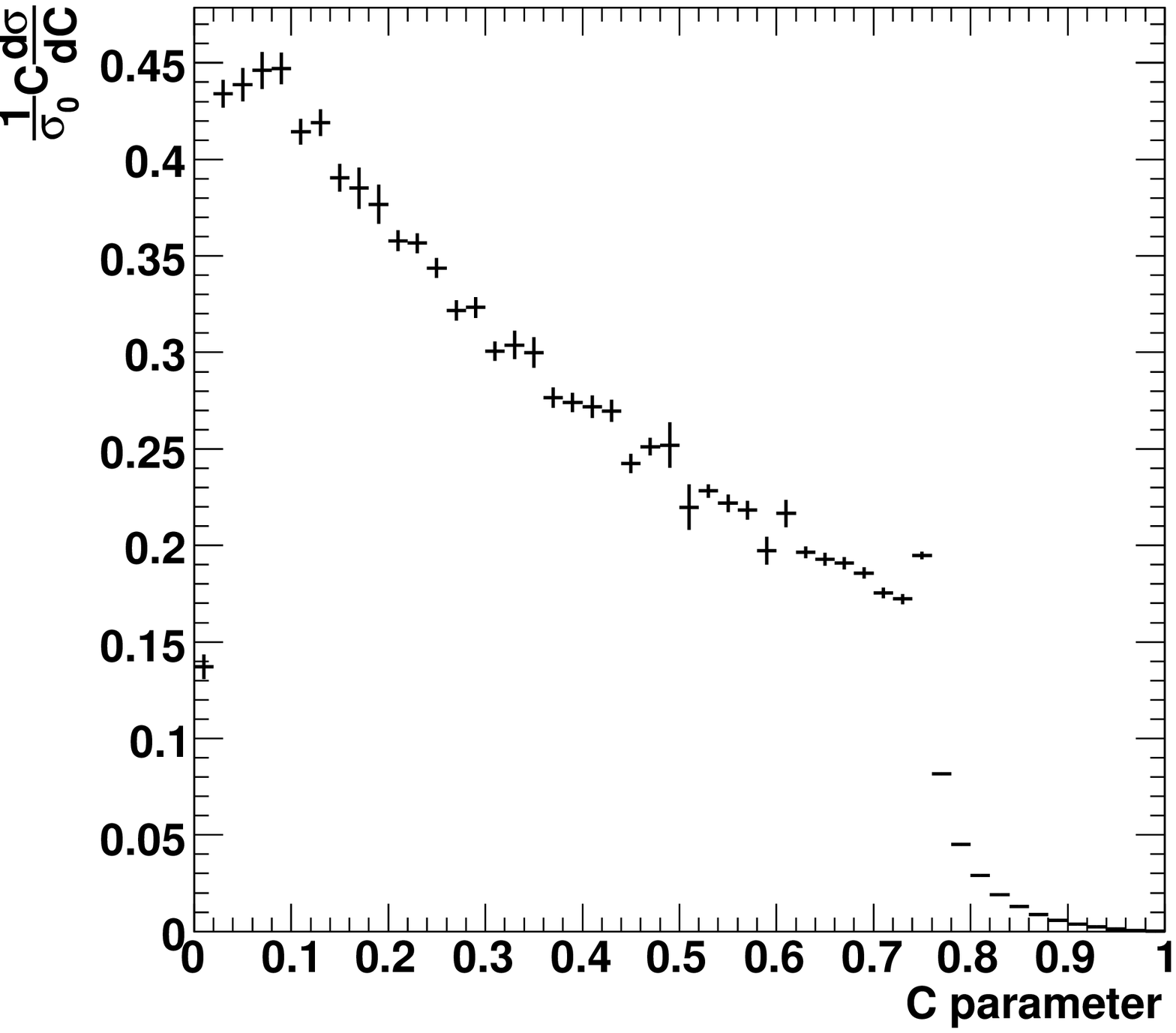}
\caption{The full next to leading order QCD prediction for the $C$-parameter.}
\label{c2}
\end{center}
\end{figure}

The differential cross section for thrust can be written, at next to leading order, as
\begin{equation}
\frac{1}{\sigma_0}(1-t)\frac{d\sigma}{dt}=\frac{\alpha_S(\mu)}{2\pi}A_t(t)+\left(\frac{\alpha_S(\mu)}{2\pi}\right)^2\left[A_t(t)2\pi b_0\log\left(\frac{\mu^2}{s}\right)+B_t(t)\right],
\end{equation}
where $\mu$ is the renormalization scale, $s$ is the square of the centre of mass energy, $\sigma_0$ is the leading order cross section for $e^+e^-\rightarrow$ hadrons, and $b_0$ is given by
\begin{equation}
b_0=\frac{33-2N_f}{12\pi}.
\end{equation}
The two coefficient functions $A_t$ and $B_t$ are universal and perturbatively calculable. Figure \ref{t1} shows the two coefficient functions, and Figure~\ref{t2} shows the full next to leading order QCD prediction (we have used $\alpha_s(M_Z)=0.118$).
The differential cross section for the $C$-parameter can be written in a similar form in terms of two coefficient functions $A_C$ and $B_C$. Figure~\ref{c1} shows these two coefficient functions, and Figure~\ref{c2} shows the full next to leading order QCD prediction for the $C$-parameter.
\subsection[$ep\rightarrow e+1$ jet]{\boldmath$ep\rightarrow e+1$ jet}
\begin{figure}
\includegraphics[width=0.5\textwidth]{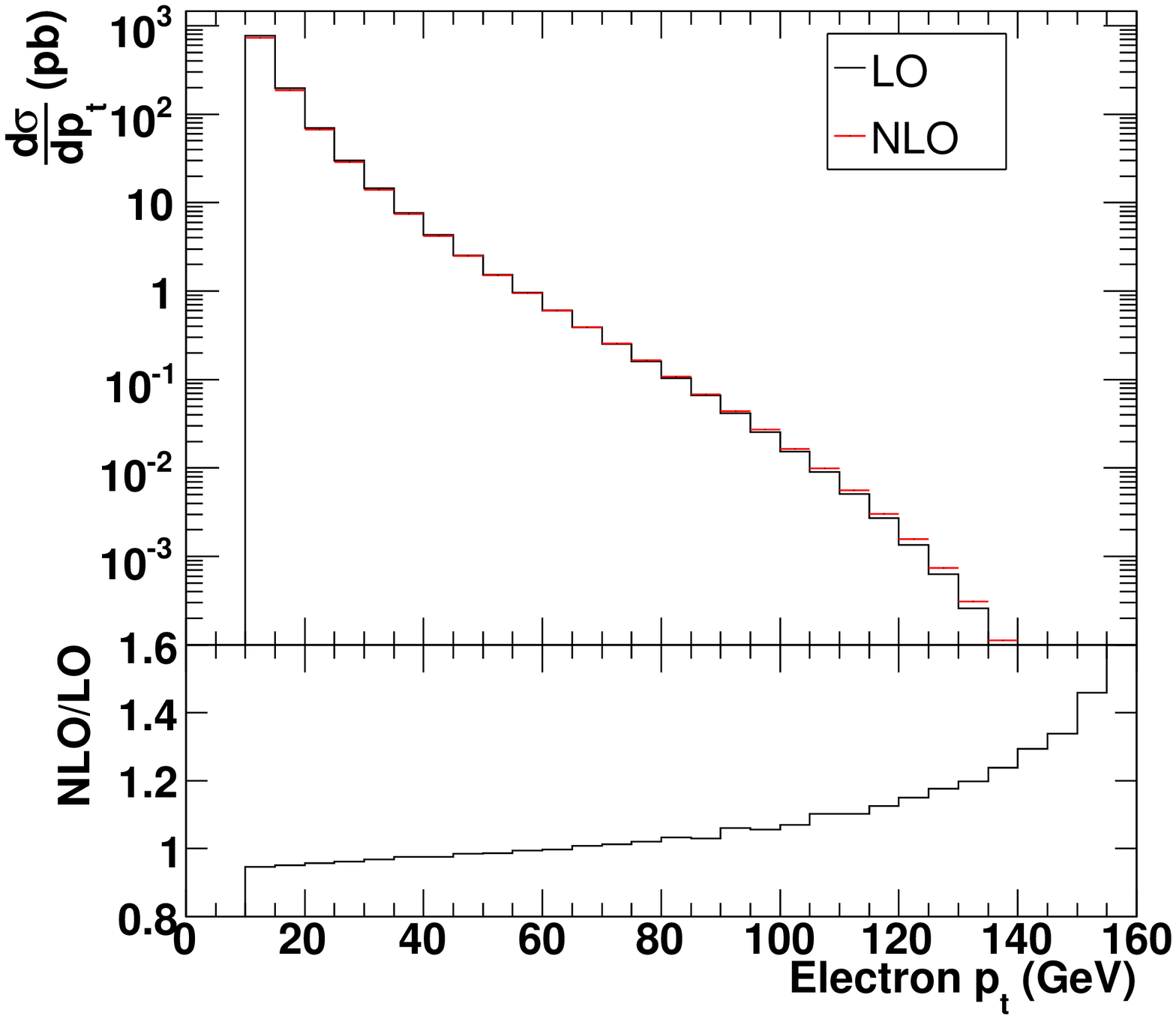}
\includegraphics[width=0.5\textwidth]{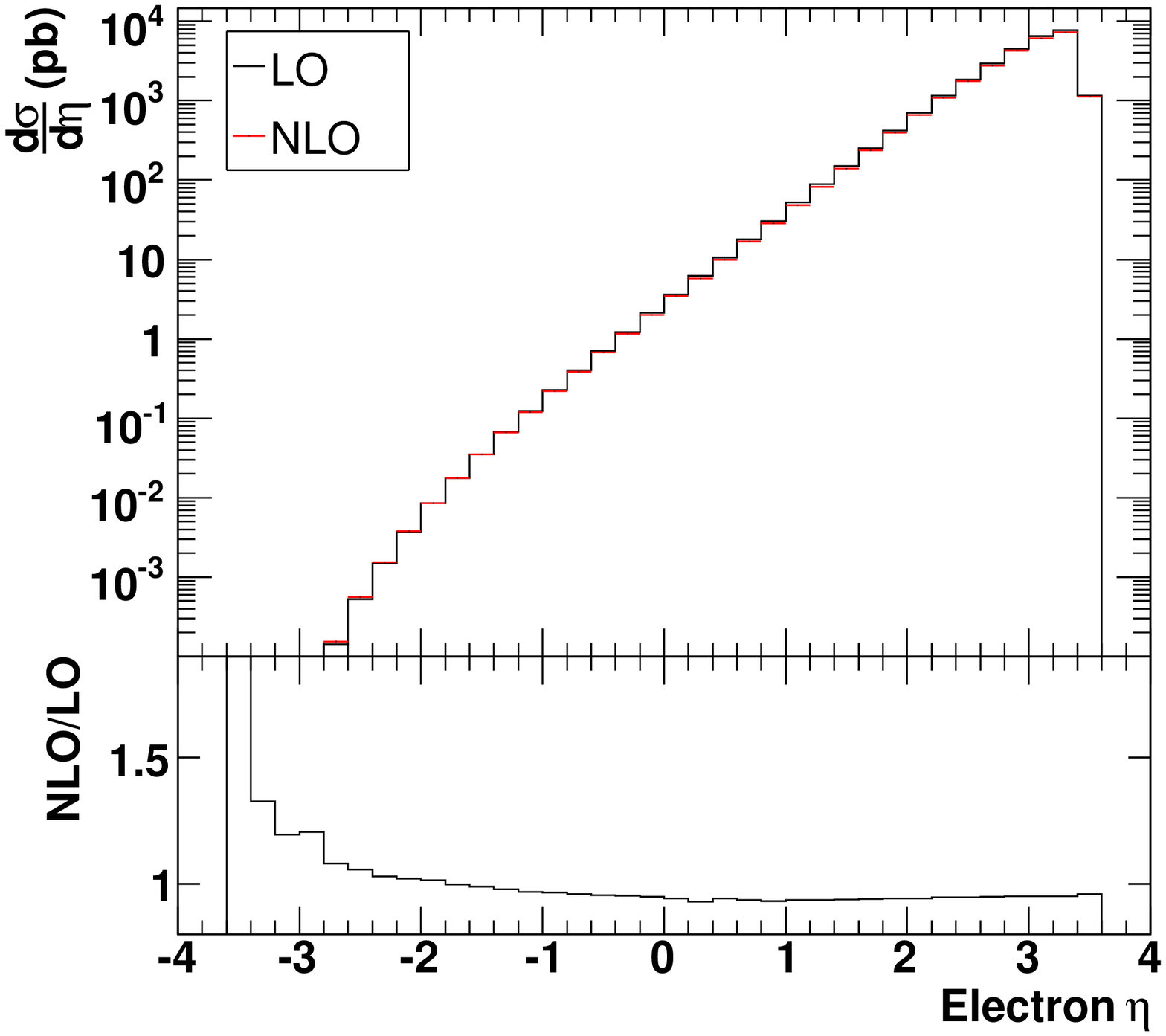}
\includegraphics[width=0.5\textwidth]{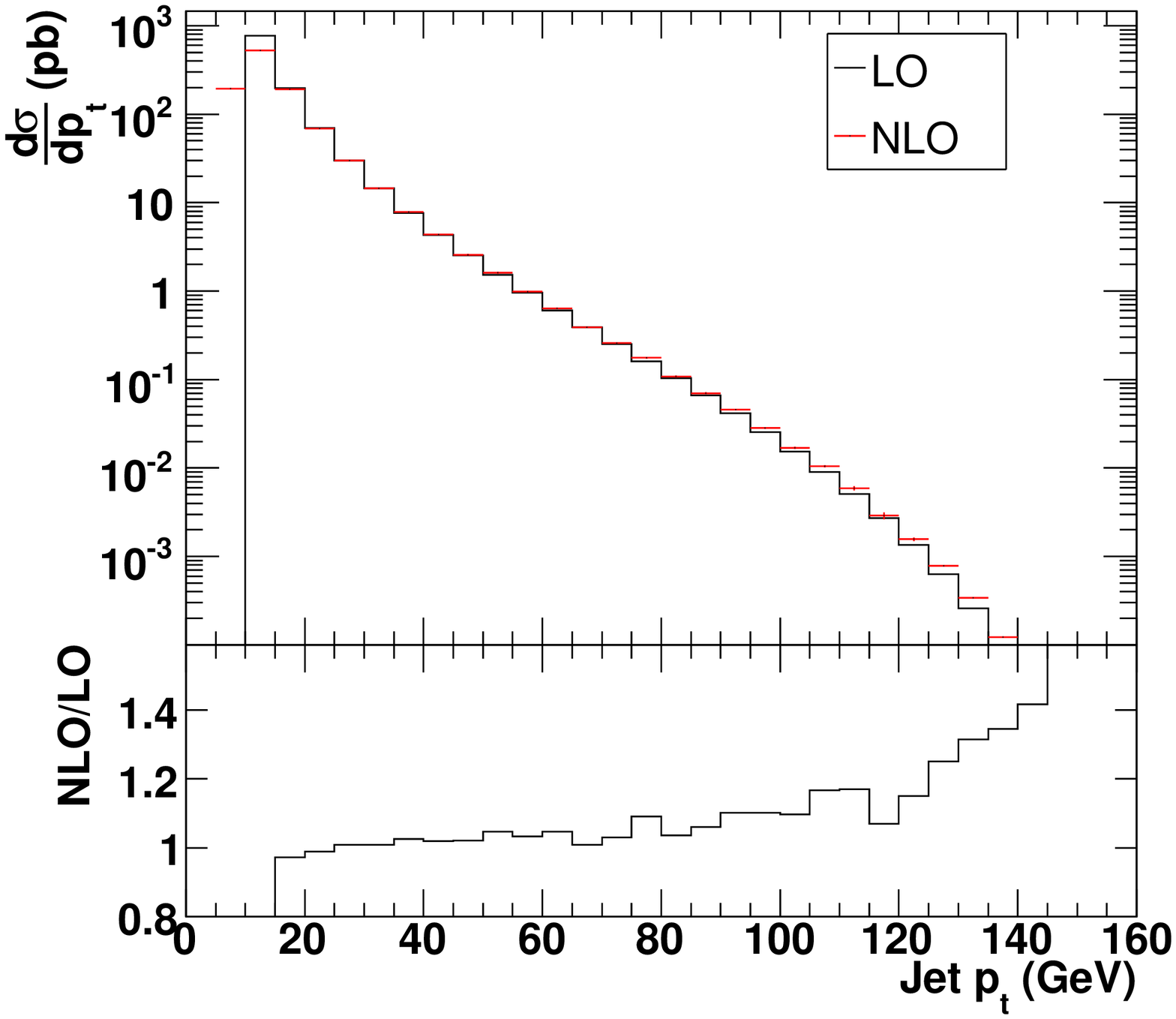}
\includegraphics[width=0.5\textwidth]{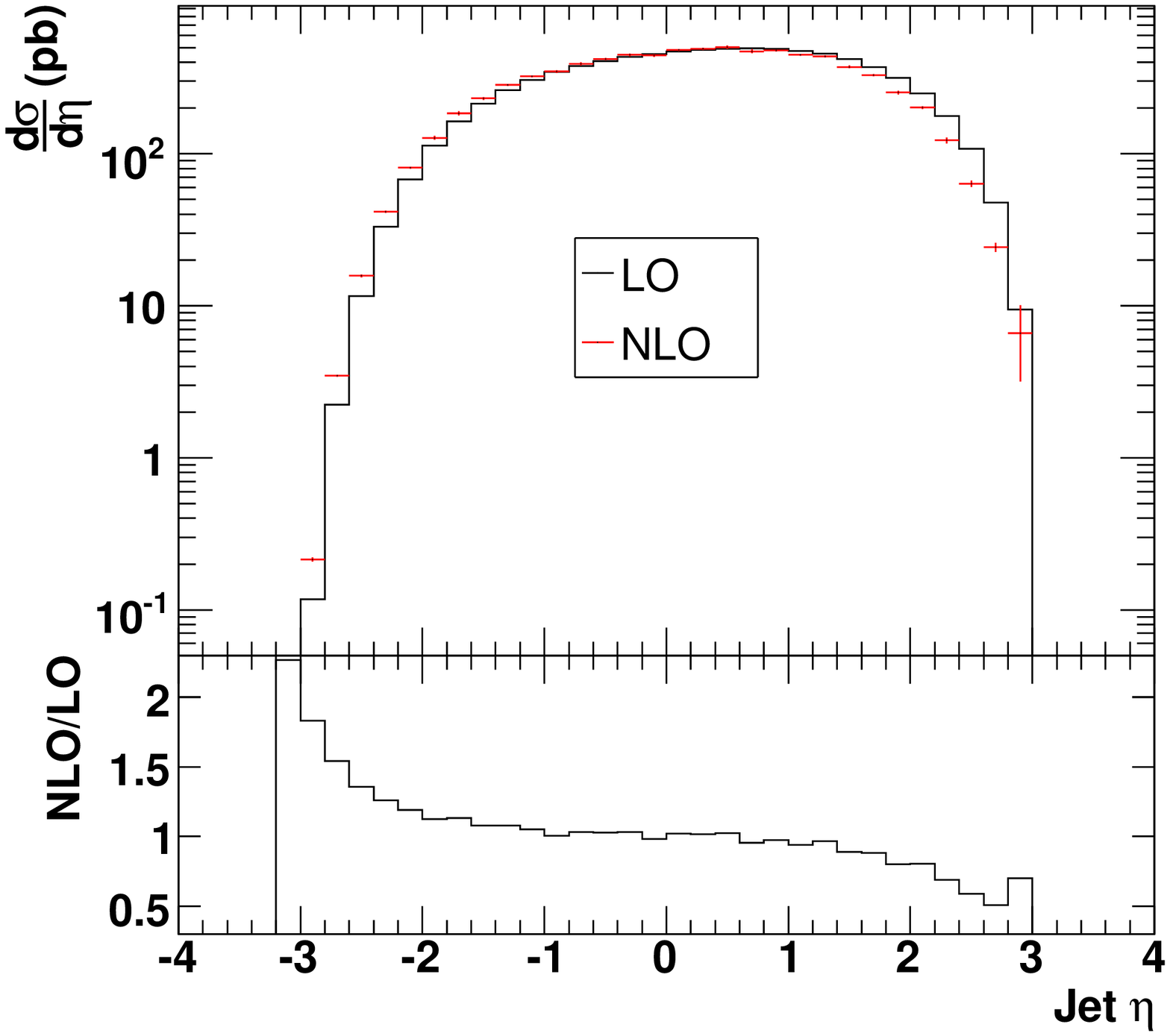}
\caption{The transverse momentum and pseudorapidity distributions of the electron and leading jet in the electron-proton centre of mass frame. A cut on the transverse momentum of the electron ($p_t>10$~GeV) has been applied, and for the pseudorapidity distribution of the leading jet a cut on the $p_t$ of the jet $p_t>15$~GeV was applied.}
\label{displots}
\end{figure}

We have also implemented the process $ep\rightarrow e+1$ jet. The beam energies chosen were a 50~GeV electron beam on a 500~GeV proton beam (i.e.\ $s=10^5$~GeV). We have used the parton distribution function CTEQ5M~\cite{Lai:1999wy} and the renormalization and factorization scales are set to $Q$ where $Q^2=-q^2$ and $q$ is the momentum transfer between the electron and the proton. A cut on the transverse momentum of the electron ($p_t>10$~GeV) was applied and to define the jets we have used the kt algorithm\cite{Catani:1993hr} as implemented in KtJet\cite{Butterworth:2002xg} in its `inclusive' mode\cite{Ellis:1993tq} with $R=1$. Figure~\ref{displots} shows the transverse momentum and pseudorapidity distributions of the electron and the leading jet in the electron-proton centre of mass frame. Good agreement with the results of Ref.~\cite{Gleisberg:2007md} is noted.
\section{Conclusion and Future Plans}
We have presented the Monte Carlo program \texttt{TeVJet} for calculating jet observables in NLO QCD, which is a direct implementation of the dipole subtraction method~\cite{Catani:1996vz}. In order to implement a new scattering process the user must provide code to calculate the matrix element squared for the Born level and real emission diagrams, and the interference between the Born and 1-loop amplitudes. In fact the user must provide the colour algebra necessary to evaluate the square of the colour correlated Born level matrix element, as well as the Born level amplitude projected onto the helicity space of each external gluon. \texttt{TeVJet} automatically calculates the subtraction terms and the operators {\bf I}, {\bf P} and {\bf K}, which are required by the dipole subtraction method; and performs the multi-channel phase space integrations and the convolution with the parton distribution functions in the case of initial state hadrons.

We have demonstrated with the example processes $e^+e^-\rightarrow 3$ jets and $ep\rightarrow e+1$ jet that the framework works and that the results agree with existing calculations. 

Our immediate plans are to extend the Monte Carlo program to include the possibility of massive external partons. Finally we note that it should be possible to fully automate the generation of code to calculate the Born level and real matrix elements so that in future the only process dependent information the user must provide is the virtual term, i.e.\ the interference between the Born and 1-loop amplitudes.

\section*{Acknowledgments}
We would like to thank Tanju Gleisberg, Andrew Pilkington and Will Plano for many useful and interesting discussions and suggestions. This work was supported in part by the UK Science and Technology Facilities Council and in part by the European Union Marie Curie Research Training Network MCnet under contract MRTN-CT-2006-035606.

\end{document}